\documentclass[12pt]{article}

\usepackage{amssymb,amsmath,amsfonts,eurosym,geometry,graphicx,caption,subcaption,color,framed,setspace,sectsty,comment,footmisc,caption,natbib,pdflscape,array,hyperref, booktabs,makecell,multirow,float,tabularx,longtable, array, url}

\setcitestyle{authoryear,round}
\usepackage[titletoc,toc,title]{appendix} 

\usepackage{authblk} 
\usepackage[left]{lineno}

\newcolumntype{L}[1]{>{\raggedright\let\newline\\arraybackslash\hspace{0pt}}m{#1}}
\newcolumntype{C}[1]{>{\centering\let\newline\\arraybackslash\hspace{0pt}}m{#1}}
\newcolumntype{R}[1]{>{\raggedleft\let\newline\\arraybackslash\hspace{0pt}}m{#1}}

\begin{document}
\modulolinenumbers[5]
\hypersetup{pageanchor=false}

\begin{titlepage}
\title{Matching-LLM: Historically Anchored Synthetic Prediction of Public Opinion and Elections\thanks{All authors have contributed equally and are listed in alphabetical order by last name. Corresponding author: Lijia Wei, ljwei@whu.edu.cn}\thanks{All data used in this article were either generated by large language models or obtained from publicly available databases. No new human participants were recruited; therefore, neither a Consent to Participate declaration nor an ethics approval number is required or applicable.}}
\author[1]{Shapeng Jiang}
\author[1]{Lijia Wei}
\author[2]{Chen Zhang}
\affil[1]{Economics and Management School, Wuhan University, China}
\affil[2]{School of Economics and Management, Huangshan University, China}
\date{}
\maketitle
\begin{abstract}
\small\setstretch{1.1}
\noindent Large language models (LLMs) offer new opportunities for predicting public opinion and electoral behavior, but their synthetic responses may not fully reflect human attitudes and heterogeneity. This study introduces Matching-LLM, a weighted framework that combines LLM-generated responses with historical human behavioral data. Using the World Values Survey, American National Election Studies, and German election survey data, we evaluate its performance in reconstructing survey responses and predicting election outcomes. Results show that Matching-LLM generally improves accuracy relative to LLM-only, including 11 of 12 comparisons on survey questions excluded from calibration, although it does not consistently outperform History-only. In the 2024 U.S. presidential election, Matching-LLM predicts 310 Republican electoral votes, compared with the actual 312, outperforming both single-source benchmarks in electoral-vote accuracy. A supplementary German application extends the framework to a multiparty setting. Overall, the findings highlight the potential of historical data to improve LLM-based predictions while emphasizing the importance of task-specific validation.
\vspace{0.2cm}

\noindent\textbf{Keywords:} Large language models; synthetic respondents; public opinion; election forecasting
\end{abstract}
\thispagestyle{empty}
\end{titlepage}
\clearpage
\hypersetup{pageanchor=true}
\setcounter{page}{1}
\doublespacing

\section{Introduction}\label{sec:introduction}

Surveys and opinion polls provide information about attitudes and choices that cannot be recovered from administrative records alone. They are essential for studying social values, trust, cultural differences, and electoral behavior, but collecting new representative responses for every question and decision environment is costly. Large language models (LLMs) offer a potentially useful supplementary source of information: a researcher can condition a model on a respondent's demographic profile and ask it to answer a questionnaire or choose among election candidates. Early studies show that such synthetic respondents can reproduce some survey patterns and experimental regularities, motivating applications in social science and economics \citep{argyle2023out,Aher2022Using,Horton2023,Dillion2023,Bail2024}. The central empirical question, however, is whether these generated responses contain reliable information about the population of interest.

The evidence is promising but uneven. Demographic conditioning can recover relationships between identities and political attitudes, yet similarity in aggregate averages does not establish fidelity to the distribution of human responses. Synthetic samples may compress variation, overstate consensus, or change substantially with wording and response format \citep{bisbee2023synthetic,Tjuatja2023Do,Rottger2024}. Work on opinion representation also shows that the perspectives expressed by a language model need not reflect the demographic and cultural composition of its intended users \citep{Santurkar2023,Durmus2024}. More recent research documents the flattening and misrepresentation of identity groups and systematic biases introduced through persona generation \citep{Wang2025,Li2025}. These findings make calibration to observed human behavior a substantive requirement, rather than a cosmetic adjustment to otherwise representative synthetic data.

Recent advances sharpen this distinction between useful predictive information and behavioral fidelity. A language model fine-tuned on large-scale human choice data can generalize across experimental tasks \citep{Binz2025}. LLM forecasts of social science experiments can correlate strongly with observed treatment effects while systematically overestimating their magnitudes \citep{Ashokkumar2026}. Agents grounded in interviews and survey self-reports can predict held-out responses more accurately than agents based only on demographics \citep{Park2026}, and new digital-twin datasets facilitate validation against rich records of actual individuals \citep{Toubia2025}. These developments support the value of human behavioral information while showing that predictive performance depends on the information supplied, the task, and the evaluation criterion. They do not imply that a general-purpose LLM, prompted with a short demographic persona, is already a calibrated substitute for a human respondent.

Historical human data provide a different source of predictive information. Earlier surveys and election outcomes reflect realized responses, including the heterogeneity and inconsistencies that synthetic respondents may omit. Social identities, regional political alignments, and cultural orientations often persist, making these records useful behavioral anchors. Research using the World Values Survey (WVS) documents both cultural change and the persistence of distinctive traditions \citep{InglehartBaker2000}. Nevertheless, historical observations are backward-looking. They do not directly incorporate a new candidate pairing, newly salient issues, changing economic conditions, or shifts in the population. Matching demographic profiles across survey waves can improve comparability, but it cannot make an old response contemporaneous or eliminate changes in the relationship between demographics and attitudes. This limitation is closely related to distributional change in prediction problems \citep{Gama2014}.

The resulting methodological problem concerns two imperfect information sources. Historical records are grounded in observed behavior but may be stale; LLM responses are conditioned on the question and scenario presented to the model but may be miscalibrated. Context conditioning means that the model can respond to the supplied candidate names and task description. It does not mean that the model automatically knows every current event or updates its training data. We ask whether these two sources can be combined to improve reconstruction and prediction, and whether an LLM contributes information beyond a historical baseline that is already strong.

We introduce \textit{Matching-LLM}, a framework that matches, aligns, and integrates historical human behavioral information with context-conditioned LLM predictions. Matching refers to bringing the two information sources into correspondence for the same target outcome and combining them at the relevant prediction level, from individual survey responses to aggregate election shares. For survey items, each target demographic profile is matched to a respondent in an earlier survey wave, and the matched historical response is combined with the LLM-generated response for that profile. For the U.S. election application, the combination operates on state-level party vote shares, using prior election results as the historical anchor. The statistical motivation comes from forecast combination: different information sources can improve prediction when their errors have suitable magnitudes and dependence, but gains are not guaranteed \citep{BatesGranger1969,Clemen1989,WangHyndman2023}. The method therefore rests on the empirical complementarity of historical and LLM-based predictions.

This contribution complements existing approaches that use human data to condition, train, or evaluate language models. For example, AI-augmented survey research combines survey information with language-model representations to predict missing opinions \citep{Kim2023AI-Augmented}. Interview-based agents and behaviorally fine-tuned models embed human information within the prediction process \citep{Park2026,Binz2025}. Matching-LLM instead provides a transparent combination at the response or aggregate-share level, with both single-source endpoints retained as explicit benchmarks. The contribution lies in integrating historically grounded human behavioral information with context-conditioned LLM predictions at different levels of aggregation, together with a comparative evaluation of when each information source adds predictive value.

Our empirical analysis proceeds through progressively more demanding validation tasks. First, we reconstruct WVS Wave 7 responses for U.S. and Chinese samples using GPT-4o and demographically matched Wave 6 observations. We examine response means, cross-country differences, dispersion, and within-person correlation patterns. Second, weights calibrated on one set of questions are applied to six additional questions that were excluded from calibration. This is cross-item transfer validation within the same survey setting. Third, we examine U.S. election calibration and a later election application, followed by a supplementary German election application in a multiparty environment. These distinctions separate fitted reconstruction, transfer to unused outcomes, and forecasting across election years, consistent with the need to distinguish prediction from explanation and to evaluate generalization directly \citep{Shmueli2010,Hofman2021}.

The results show substantial gains relative to LLM-only responses, alongside clear limits to the incremental value of the LLM signal. Historical anchoring improves many WVS mean comparisons and restores variation where the LLM generates nearly constant answers. In the U.S., Matching-LLM reduces the average correlation gap relative to History-only; the corresponding improvement does not occur in China. On the six unused survey questions, Matching-LLM is closer to the human mean than LLM-only in eleven of twelve country--item comparisons, but History-only is closer than the combination on all six Chinese items. The appropriate conclusion is therefore task-dependent complementarity, rather than uniform dominance of either information source.

For the 2024 U.S. application, a historical weight of $h=0.82$ produces 310 Republican electoral votes, compared with the actual 312, the LLM-only prediction of 319, and the History-only prediction of 291. History-only and Matching-LLM each misclassify two states, although the combination improves the electoral-vote total and the composition of correctly predicted swing states. The weight is obtained from continuous mean squared error (MSE) calibration using the 2020 outcome. The full-sample MSE-minimizing weight is 0.8990, with a 95\% bootstrap interval of approximately $[0.8203,0.9698]$; its lower endpoint, rounded to two decimal places, gives the application weight of 0.82. The interval summarizes the sensitivity of the estimated historical weight to the composition of the state-level calibration sample. The German application extends Matching-LLM to a different electoral system through party-specific historical calibration and presents a supplementary comparison of predicted and observed outcomes.

The paper makes three connected contributions. It develops a historically anchored synthetic prediction framework; evaluates that framework beyond mean similarity, using cross-cultural contrasts, response variation, and correlation structure; and tests whether historical information and LLM output offer incremental value in later election applications. These contributions position synthetic responses as a supplement to empirical measurement that must be benchmarked against human data, consistent with guidance on the use of silicon samples \citep{Sarstedt2024}. Section~\ref{sec:design} presents the data and methodology. Section~\ref{sec:result} evaluates survey reconstruction and cross-item transfer. Section~\ref{sec:elections} reports election applications, Section~\ref{sec:discussion} discusses their interpretation and limitations, and Section~\ref{sec:conclusion} concludes.

\section{Data and Methodology}\label{sec:design}

\subsection{Data Sources and Prediction Tasks}\label{subsec:data}

The main survey analysis uses WVS Waves 6 and 7, which collect repeated cross-sectional information on values and attitudes across countries \citep{WVS6,WVS7}. The Wave 7 analytical samples contain 2{,}077 U.S. respondents and 2{,}001 Chinese respondents. Wave 7 demographic profiles define the personas supplied to the LLM, and actual Wave 7 answers provide the human benchmark. Wave 6 supplies historical responses to questions observed in both waves. Because different individuals participate in the two waves, matched records are comparable demographic profiles rather than repeated observations of the same person.

The calibration set contains fourteen common items: three social-value items, four trust items, and seven ethical-norm items. Social-value items use harmonized 3-point scales, trust items use 4-point scales, and ethical-norm items use 10-point scales. Six additional 4-point items concerning gender roles and family values are excluded from weight estimation and used for cross-item validation. The numerical coding of an item must be read together with its wording; a higher value does not have the same substantive meaning across all question blocks. Supplementary Appendix A describes demographic variables, and Appendix C contains the full prompts.

The main election application uses demographic profiles from the American National Election Studies (ANES) 2020 Time Series Study \citep{ANES2020}. The analytical sample used in the simulations contains 6{,}571 respondents. The same profiles are used to generate synthetic U.S. presidential votes, which are aggregated using survey sampling weights. Historical state-level party vote shares provide a separate behavioral benchmark. The 2020 election is the calibration target, and the combination is then applied to the 2024 Trump--Harris contest. The German application, described in Section~\ref{subsec:germany}, is a supplementary test using GLES profiles and party-specific multiplicative calibration coefficients.

No new human participants were recruited for this study. The analysis uses existing survey records and LLM-generated responses, and none of the studies was preregistered. Observed Wave 7 outcomes enter WVS calibration, while the election applications use LLM simulations generated before the corresponding elections. The empirical design distinguishes survey reconstruction, transfer to unused items, and prediction of subsequent election outcomes.

\subsection{Synthetic Responses and Historical Baselines}\label{subsec:signals}

All synthetic responses are generated through the GPT-4o API, without a real-time search component. A demographic persona specifies the respondent's observable characteristics, and the model is asked to answer the relevant survey question or select among the named candidates or parties. The full system and user prompts remain in Supplementary Appendix C. The prompts supply context to the generation process; the method does not infer unobserved personal histories from actual respondents or assume that the LLM retrieves contemporaneous news. We refer to the resulting predictions as \textit{LLM-only}, or $L$.

For WVS, generation is conditioned on the Wave 7 persona, while the historical signal is constructed separately. We use within-country 1:1 nearest-neighbor propensity score matching to associate each Wave 7 profile with a Wave 6 profile, using age, gender, education, marital status, occupation, and income scale. The matched Wave 6 answer to each common question defines \textit{History-only}, or $H$. The same matched historical records are used in History-only and Matching-LLM, so the comparison isolates the contribution of adding the generated signal to that historical baseline.

Demographic propensity score matching addresses observed compositional differences, but not all differences between waves. Individuals with similar demographics can have different attitudes, and historical donors may not represent current respondents' unobserved experiences. The high quality of demographic correspondence is important for constructing a useful historical signal, but it does not proof that a historical donor is the target respondent's behavioral equivalent \citep{Stuart2010}.

For the U.S. election application, the historical anchor is defined at the state--party level. Let $v_{s,t,p}$ denote the actual vote share of party $p$ in state $s$ and election year $t$. The two historical baselines are
\begin{equation}
 H_{s,2020,p}=\frac{v_{s,2012,p}+v_{s,2016,p}}{2},\qquad
 H_{s,2024,p}=\frac{v_{s,2016,p}+v_{s,2020,p}}{2}.
 \label{eq:historical-votes}
\end{equation}
Thus, the 2020 outcome is used to evaluate the 2020 calibration, but does not enter its own historical predictor. The 2024 baseline uses only the two preceding elections. Individual responses from the historical WVS wave and the aggregation of state-level historical election outcomes in the U.S. application represent historical-signal construction at two different levels. In both applications, Matching-LLM subsequently aligns and combines the historical and LLM-based signals at the corresponding prediction level.

Let $z_{i,t,p}$ indicate whether persona $i$ votes for party $p$ in the simulation, $w_i$ be the survey weight, and $\mathcal I_s$ be the respondent set used for state $s$. The LLM state-level share is the weighted aggregation
\begin{equation}
 L_{s,t,p}=\frac{\sum_{i\in\mathcal I_s}w_i z_{i,t,p}}
 {\sum_{i\in\mathcal I_s}w_i}.
 \label{eq:llm-votes}
\end{equation}
The 2024 simulations use demographic profiles from ANES 2020 rather than a new sample of 2024 voters. Although ANES survey weights are applied when aggregating simulated choices, the resulting state-level estimates may still be affected by errors in LLM-generated votes, small state subsamples, and changes in the electorate between 2020 and 2024.

\subsection{The Matching-LLM Framework}\label{subsec:framework}

Matching-LLM aligns and integrates historical human behavioral information and generated predictions of the \emph{same} target. At the individual--item level in WVS, or the state--party level in the U.S. application, the combination is
\begin{equation}
 M(h)=hH+(1-h)L=L+h(H-L),\qquad 0\leq h\leq1.
 \label{eq:combination}
\end{equation}
Here $h$ is the historical weight. The endpoints $h=0$ and $h=1$ reproduce LLM-only and History-only, respectively. At an interior weight the historical signal adjusts the generated response toward the observed behavioral anchor. Equivalently, the LLM signal adjusts the historical baseline toward the response implied by the supplied context.

The motivation is that some behavior persists across settings, while other behavior changes with the question, candidates, and social environment. Historical observations can inform persistence but also contain outdated context; LLM output can respond to the current task description but also contain persistent model bias. We do not want to separately identify stable behavioral patterns from historical data or context-specific responses from LLM predictions. Instead, both sources are treated as predictions of the same outcome and combined through the weighted framework.

Forecast-combination theory provides a direct statistical account of the possible gains \citep{BatesGranger1969,Clemen1989,WangHyndman2023}. Let $Y$ be the target, and define $e_H=H-Y$ and $e_L=L-Y$. If their means are $b_H,b_L$, variances are $\sigma_H^2,\sigma_L^2$, and covariance is $c$, then
\begin{align}
 \operatorname{MSE}\{M(h)\}
 &=\{hb_H+(1-h)b_L\}^{2}\nonumber\\
 &\quad+h^{2}\sigma_H^{2}+(1-h)^{2}\sigma_L^{2}
       +2h(1-h)c.
 \label{eq:mse}
\end{align}
Combination performance depends jointly on systematic bias ($b_H,b_L$), dispersion of errors ($\sigma_H^2,\sigma_L^2$), and their dependence ($c=\operatorname{Cov}(e_H,e_L)$). Errors with different directions can partly offset, but two signals sharing a large bias can remain inaccurate even after combination. Imperfect correlation alone is not sufficient to guarantee improvement over the better single source.

To make this condition explicit, let $A=\mathbb E(e_H^2)$, $B=\mathbb E(e_L^2)$, and $C=\mathbb E(e_He_L)$. When $A+B-2C>0$, the population MSE-optimal convex weight is
\begin{equation}
 h^*=\operatorname{clip}_{[0,1]}
 \left(\frac{B-C}{A+B-2C}\right),
 \label{eq:optimal-weight}
\end{equation}
where $\operatorname{clip}_{[0,1]}$ restricts the value to the unit interval. An interior optimum requires $C<\min(A,B)$ and then improves population MSE relative to both endpoints. If this condition fails, the best convex combination can be a single-source endpoint. The MSE formulation illustrates how complementary prediction errors can improve a weighted forecast. Both WVS and U.S. election calibration minimize MSE. WVS calibration evaluates individual responses, whereas U.S. election calibration evaluates state-level party vote shares. The two applications therefore share a quadratic calibration loss while differing in the level of aggregation and the construction of historical information.

It is important to note that LLM-only may capture broad response patterns while possibly missing dispersion or calibration. History-only may perform well when behavioral relationships persist, yet fail under contextual change. An interior combination is useful only where its observed losses improve on an appropriate baseline. Finally, the informative weight can differ across tasks and countries as error structures change, and we evaluate these expectations through the available contrasts. A large $h$ indicates that the selected prediction relies heavily on history, but it is not a numerical measure of behavioral persistence independent of the model and loss function.

\subsection{Weight Calibration and Evaluation Strategy}\label{subsec:calibration}

\noindent\textbf{WVS calibration and transfer.} For country $c$, let $\mathbf R_{c,k}^{H}$, $\mathbf R_{c,k}^{L}$, and $\mathbf R_{c,k}^{Y}$ be historical, generated, and actual Wave 7 response vectors for question $k$, each containing $N_c$ individual responses. A single country-specific weight minimizes the mean squared error across respondents and the fourteen calibration items $\mathcal K_{\mathrm{cal}}$:
\begin{equation}
 \widehat h_c=\arg\min_{h\in[0,1]}
 \frac{1}{N_c|\mathcal K_{\mathrm{cal}}|}
 \sum_{k\in\mathcal K_{\mathrm{cal}}}
 \left\|h\mathbf R_{c,k}^{H}+(1-h)\mathbf R_{c,k}^{L}
             -\mathbf R_{c,k}^{Y}\right\|_2^{2}.
 \label{eq:wvs-calibration}
\end{equation}
Separate calibration yields $\widehat h_{\mathrm{US}}=0.23$ and $\widehat h_{\mathrm{China}}=0.31$. Weighted survey responses can take fractional values, which we interpret as numerical response indices on the original coding scale.

The calibrated combination minimizes aggregate MSE over a set that includes both single-source predictors. Its relative performance on individual items and other evaluation measures is assessed empirically. The fourteen-item comparisons evaluate reconstruction and calibration in the observed survey setting. We then hold the country-specific weights fixed for six questions excluded from $\mathcal K_{\mathrm{cal}}$, assessing cross-item transfer within the same respondents and survey period.

\noindent\textbf{U.S. election calibration.} We estimate the historical weight by minimizing state-level party vote-share MSE over the continuous interval $h\in[0,1]$, using actual 2020 outcomes as calibration targets. The criterion averages squared errors over Democratic and Republican shares in the fifty states and the District of Columbia, giving equal weight to each jurisdiction. The primary loss is
\begin{equation}
 \operatorname{MSE}_{2020}(h)=\frac{1}{2S}
 \sum_{s=1}^{S}\sum_{p\in\{D,R\}}
 \left[hH_{s,2020,p}+(1-h)L_{s,2020,p}-v_{s,2020,p}\right]^2,
 \quad S=51.
 \label{eq:election-mse}
\end{equation}
The MSE criterion is a continuous convex quadratic function of $h$, and its constrained minimum can be obtained analytically. The estimated weight is
\[
 \widehat h_{\mathrm{MSE}}=\arg\min_{h\in[0,1]}\operatorname{MSE}_{2020}(h).
\]
The full-sample optimum is $\widehat h_{\mathrm{MSE}}=0.8990$. The dense sampling of historical weights used in the sensitivity figure serves only to display the continuous loss curve and does not determine the estimated optimum.

To assess sensitivity to the composition of the state-level calibration sample, we use 10{,}000 bootstrap replications, following the resampling principle of \citet{Efron1979}. Each replication samples 51 jurisdictions with replacement, retains the Democratic and Republican observations together for each sampled jurisdiction, and re-estimates the continuous MSE-minimizing weight. Historical shares, LLM-generated shares, and observed 2020 shares are held fixed within each jurisdiction. The 2.5th and 97.5th percentiles of the resulting estimates give a 95\% interval of approximately $[0.8203,0.9698]$. This interval summarizes weight sensitivity to state-level sample composition, conditional on the available calibration inputs.

For the 2024 forecast, we use the lower endpoint rounded to two decimal places, $h=0.82$, assigning 82\% weight to the historical signal and 18\% to the LLM signal. This choice retains a strong historical anchor while giving context-conditioned information greater influence than under the full-sample MSE optimum. The historical inputs are advanced from 2012--2016 to 2016--2020, while the application weight is held fixed. Actual 2024 outcomes do not enter the calibration.

Vote-share MSE is calculated from shares expressed as proportions between zero and one. For presentation, it is multiplied by $100^2$ and reported in squared percentage points.

\noindent\textbf{Evaluation dimensions.} For survey items we compare response means and standard deviations, U.S.--China mean differences, and Pearson correlations within question blocks. Table~\ref{tab:enhanced} reports absolute deviations of response means, denoted MAD, together with two-sided t-tests comparing the mean deviations of LLM-only and Matching-LLM from the human benchmark. These are distinct from individual-level mean absolute prediction errors. Figure~\ref{fig:mean-sd} and Figure~\ref{fig:new-qustions-WVS} use standard-deviation bars, whereas Figure~\ref{fig:us-cn-diff} uses 95\% confidence intervals. Neither small synthetic standard errors nor visual overlap with a human standard deviation establishes behavioral equivalence \citep{Sarstedt2024}.

For elections we distinguish vote-share MSE, the number of correct state winners, electoral-vote error, and correctness of the national winner. These measures assess the accuracy of party support, geographic voting patterns, the aggregate electoral outcome, and the predicted winner. These applications report point predictions, not calibrated outcome probabilities; proper scoring rules would be appropriate for a future probabilistic extension \citep{GneitingRaftery2007}. Throughout, the relevant incremental comparison includes History-only as well as LLM-only.

\section{Survey Response Reconstruction and Validation}\label{sec:result}

\subsection{Response-Level Calibration}\label{subsec:response-calibration}

\begin{figure}[!ht]
    \centering
    \includegraphics[width=1\linewidth]{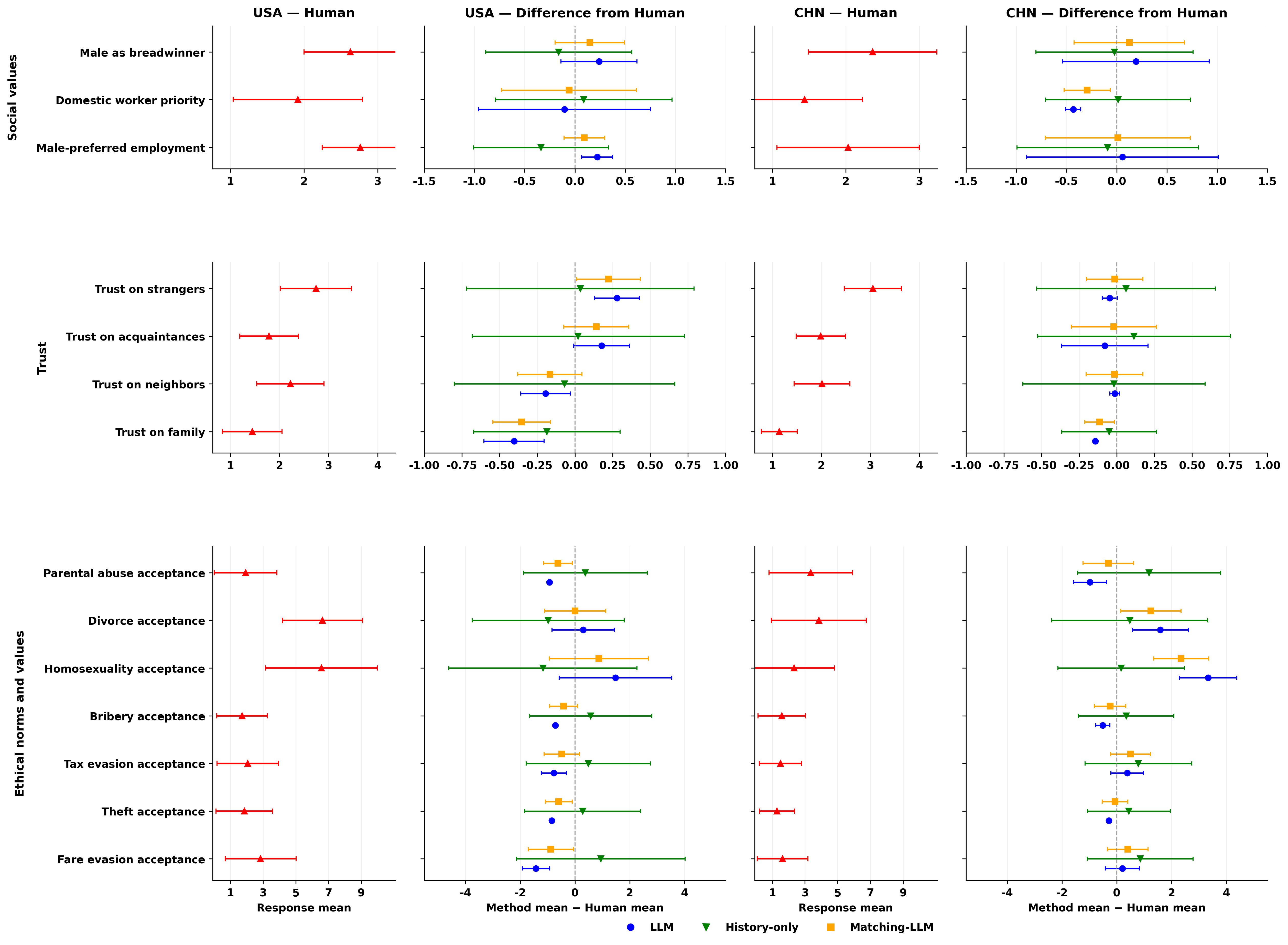}
    \caption{Human response means and method deviations from the human benchmark.\label{fig:mean-sd}}
    \caption*{\footnotesize \setstretch{1} \textit{Notes: }The figure has four columns: the U.S. human benchmark, U.S. method deviations, the China human benchmark, and China method deviations. Red upward triangles show WVS Wave 7 human response means, with horizontal bars extending one standard deviation on either side. In the deviation columns, blue circles denote LLM-only, green downward triangles denote History-only, and orange squares denote Matching-LLM. Each point is the method mean minus the corresponding Wave 7 human mean; zero therefore indicates agreement in means. Horizontal bars in these columns show the respective method's standard deviation around the displayed mean deviation. History-only uses demographically matched Wave 6 responses; Matching-LLM combines those responses with LLM-generated Wave 7 responses using historical weights of 0.23 for the U.S. ($N=2{,}077$) and 0.31 for China ($N=2{,}001$). Rows are grouped into social values (3-point harmonized scales), trust (4-point scales), and ethical norms and values (10-point scales). Demographic definitions and full question prompts are provided in Supplementary Appendices A and C.}
\end{figure}

Figure~\ref{fig:mean-sd} compares actual WVS Wave 7 responses with LLM-only, History-only, and Matching-LLM. For each country, one column reports human means and the adjacent column reports deviations of each method's mean from that benchmark. The historical weights are 0.23 for the U.S. and 0.31 for China. The fourteen items are used in weight calibration, and these comparisons present the resulting reconstruction and calibration of responses in the observed survey setting. Beyond these calibration results, subsequent analyses examine cross-item transfer using survey questions excluded from calibration and evaluate forecasting performance in subsequent election applications.

The LLM captures some broad response patterns, but its errors vary substantially by item and country. On several ethical-norm questions, generated responses are concentrated near the least accepting category. Theft and bribery responses in the U.S. are constant at the scale minimum (the lowest coded category, indicating that the behavior is never justifiable), while parental-abuse responses show very little variation. These outputs contrast with the dispersion in actual human answers. Although these generated answers are close to a seemingly plausible normative response, their limited variation prevents them from reproducing the heterogeneity of the human response distribution. This pattern is consistent with the broader evidence that aggregate similarity can coexist with synthetic homogenization \citep{bisbee2023synthetic,Wang2025}.

Value-sensitive items also reveal substantial errors in response levels. On homosexuality acceptance in China, the human mean is approximately 2.341, the LLM-only mean is 5.675, and the Matching-LLM mean is 4.691. Historical anchoring reduces the deviation but leaves a large discrepancy; the History-only mean of 2.501 is substantially closer to the benchmark. In the U.S., the corresponding human, generated, and combined means are approximately 6.565, 8.047, and 7.438. These differences are compatible with uneven cultural representation and value-related model preferences documented elsewhere \citep{Durmus2024,Santurkar2023,scherrer2024evaluating}.

The comparison also provides examples of complementary mean errors. For U.S. divorce acceptance, LLM-only is above the human mean and History-only is below it; the combination is close to the benchmark, with means of approximately 6.936, 5.661, and 6.643 compared with the human 6.636. This is the pattern anticipated by the error-combination framework. By contrast, the historical and LLM signals both overstate Chinese acceptance of fare evasion and tax evasion, and the combination moves farther from the benchmark than LLM-only. Historical grounding can therefore mitigate an error or reinforce it, depending on its direction and magnitude.

Table~\ref{tab:enhanced} reports the absolute deviations of the LLM-only and Matching-LLM means. The combined response mean is closer to the human mean for all fourteen U.S. calibration items. Examples include reductions from 0.222 to 0.093 for male-preferred employment, from 1.482 to 0.873 for homosexuality acceptance, and from 0.301 to 0.007 for divorce acceptance. In China, eleven of fourteen deviations decline. Fare evasion and tax evasion show larger increases, while trust in neighbors increases slightly from 0.014 to 0.015. The reported significance annotations are not uniform: male-preferred employment and trust in neighbors in China do not show statistically significant differences. These results support improved calibration relative to LLM-only on many items.

\begin{table}[!ht]
\def\sym#1{\ifmmode^{#1}\else\(^{#1}\)\fi}
\footnotesize
\centering
\caption{Absolute deviations of LLM-only and Matching-LLM response means. \label{tab:enhanced}}
\begin{tabular}{@{}lcccccc@{}}
\toprule
        & \multicolumn{3}{c}{U.S.} & \multicolumn{3}{c}{China} \\ 
        \cmidrule(l){2-4} \cmidrule(l){5-7}
& LLM & Matching-LLM &  & LLM & Matching-LLM &  \\
Survey Question & MAD & MAD & Difference & MAD & MAD & Difference \\
    &(1) &(2) &(3) &(4) &(5) &(6) \\ 
\midrule
\multicolumn{7}{l}{\textbf{Panel A: Social values}} \\
Male-preferred employment & 0.222 & 0.093 & 0.129\sym{***} & 0.056 & 0.010 & 0.046 \\
Domestic worker priority & 0.103 & 0.059 & 0.044\sym{*} & 0.434 & 0.295 & 0.139\sym{***} \\
Male as breadwinner & 0.240 & 0.148 & 0.092\sym{***} & 0.190 & 0.124 & 0.066\sym{**} \\
\midrule
\multicolumn{7}{l}{\textbf{Panel B: Trust}}            \\
Trust on family & 0.404 & 0.354 & 0.050\sym{***} & 0.143 & 0.114 & 0.029\sym{***} \\
Trust on neighbors & 0.195 & 0.166 & 0.029\sym{*} & 0.014 & 0.015 & 0.001 \\
Trust on acquaintances & 0.178 & 0.142 & 0.036\sym{**} & 0.079 & 0.019 & 0.060\sym{***} \\
Trust on strangers & 0.279 & 0.223 & 0.056\sym{***} & 0.047 & 0.013 & 0.034\sym{*} \\
\midrule
\multicolumn{7}{l}{\textbf{Panel C: Ethical norms and values}}            \\
Fare evasion acceptance & 1.425 & 0.88 & 0.545\sym{***} & 0.202 & 0.406 & 0.203\sym{***} \\
Theft acceptance & 0.851 & 0.59 & 0.261\sym{***} & 0.292 & 0.065 & 0.227\sym{***} \\
Tax evasion acceptance & 0.770 & 0.481 & 0.289\sym{***} & 0.380 & 0.507 & 0.127\sym{***} \\
Bribery acceptance & 0.717 & 0.420 & 0.297\sym{***} & 0.511 & 0.246 & 0.265\sym{***} \\
Homosexuality acceptance & 1.482 & 0.873 & 0.609\sym{***} & 3.334 & 2.350 & 0.984\sym{***} \\
Divorce acceptance & 0.301 & 0.007 & 0.294\sym{***} & 1.591 & 1.245 & 0.346\sym{***} \\
Parental abuse acceptance & 0.926 & 0.625 & 0.301\sym{***} & 0.977 & 0.308 & 0.669\sym{***} \\
\bottomrule
\end{tabular}
    \vspace{0.3cm} 
    \par\raggedright 
    \small \textit{Notes:} MAD denotes the absolute deviation of a method's response mean from the corresponding WVS Wave 7 human mean, measured in the original item scale. Columns (1) and (4) report LLM-only deviations; columns (2) and (5) report Matching-LLM deviations. The annotations report two-sided t-tests; *, **, and *** denote the 10\%, 5\%, and 1\% levels. Direction of improvement should be determined by comparing the MAD columns directly: some displayed Difference entries are unsigned magnitudes. The historical weights, 0.23 for the United States and 0.31 for China, are calibrated on these fourteen items. The human analytical samples contain 2{,}077 U.S. and 2{,}001 Chinese respondents. History-only comparisons can be seen in Figure~\ref{fig:mean-sd}.
\end{table}

Figure~\ref{fig:mean-sd} further illustrates the value of historical information in improving LLM-generated responses. For several trust items and some Chinese social-value and ethical-norm items, History-only closely approximates the observed human means, highlighting the predictive value of historical behavioral data. By incorporating this information, Matching-LLM improves substantially on LLM-only across most calibration items and, in some cases, also outperforms History-only. Although the relative performance of the combined approach varies across questions, these findings suggest that historical anchoring provides an effective means of improving LLM-based survey predictions, with the additional contribution of LLM information depending on the specific outcome.

Beyond mean accuracy, Figure~\ref{fig:mean-sd} also illustrates the contribution of historical information to response heterogeneity. Consistent with previous findings that LLM-generated responses may exhibit insufficient variation \citep{bisbee2023synthetic,Wang2025}, LLM-only produces highly concentrated responses for several ethical-norm items. Incorporating matched historical responses introduces additional variation into the synthetic data, particularly for items where LLM-only responses are nearly constant. Although the resulting distributions do not always fully reproduce human response variability, historical anchoring helps address a recognized limitation of LLM-based survey simulation.

Overall, the results demonstrate the effectiveness of Matching-LLM in improving the reconstruction of survey responses. Relative to LLM-only, the combined approach reduces mean deviations for all fourteen calibration items in the United States and eleven of fourteen items in China, while also mitigating the limited response variation observed in several questions. These findings suggest that integrating historical human responses with LLM-generated predictions offers a practical approach to improving the accuracy and heterogeneity of synthetic survey data. Building on these results, the following sections examine whether these improvements extend to cross-cultural differences, correlation structures, and survey questions excluded from weight calibration.

\subsection{Cross-Cultural Differences}\label{subsec:cultural-differences}

\begin{figure}[!ht]
    \centering
    \includegraphics[width=1\linewidth]{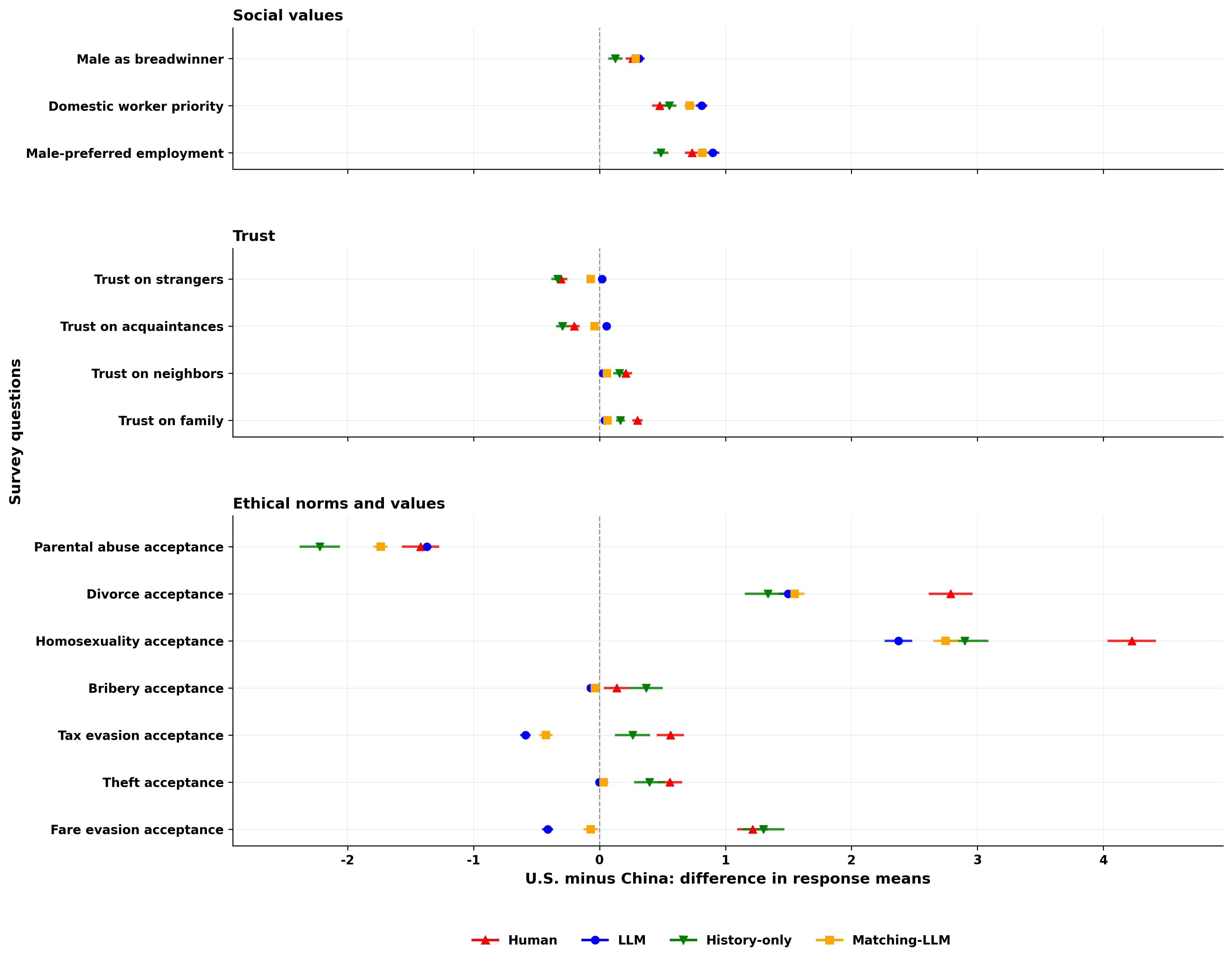}
    \caption{U.S.--China differences in response means.\label{fig:us-cn-diff}}
    \caption*{\footnotesize \setstretch{1} \textit{Notes: }Each point is the U.S. response mean minus the China response mean for the indicated item. Red upward triangles denote WVS Wave 7 human responses, blue circles LLM-only, green downward triangles History-only, and orange squares Matching-LLM. Horizontal bars are 95\% confidence intervals for the differences in means. The vertical dashed line marks zero. History-only uses the matched WVS Wave 6 responses; Matching-LLM assigns historical weights of 0.23 for the U.S. and 0.31 for China. Items are grouped into social values, trust, and ethical norms and values. The Wave 7 samples contain 2{,}077 U.S. and 2{,}001 Chinese respondents.}
\end{figure}

Figure~\ref{fig:us-cn-diff} evaluates a distinct target: the difference between the U.S. and Chinese response means for each item. A method can reproduce the within-country ordering of answers yet misrepresent the distance between countries. The figure therefore reports both the sign and magnitude of each U.S.--China contrast, with 95\% confidence intervals around the estimated differences. These comparisons capture differences in average responses between the U.S. and Chinese survey samples.

All three methods capture the positive contrasts for the displayed social-value items, but their magnitudes differ. For domestic worker priority, the human contrast is approximately 0.478, compared with 0.810 for LLM-only, 0.553 for History-only, and 0.715 for Matching-LLM. For male-preferred employment, the corresponding values are 0.734, 0.900, 0.487, and 0.816. The combination moves the generated contrasts toward the benchmark.

The trust items provide a particularly clear sign correction. Human U.S.--China differences are negative for trust in acquaintances and strangers, whereas the LLM produces positive differences. Matching-LLM restores the negative sign, using historical contrasts that are also negative. Yet the combined differences remain much smaller in absolute magnitude than the human contrasts. For trust in strangers, the human difference is approximately $-0.306$, compared with $0.020$ for LLM-only, $-0.332$ for History-only, and $-0.069$ for Matching-LLM. Restoring the direction of the contrast represents an improvement over LLM-only, even though the estimated gap remains small.

The ethical-norm items provide further evidence of how historical information can improve cross-country comparisons. For theft acceptance, LLM-only captures little of the positive U.S.--China difference observed in human responses, whereas Matching-LLM recovers its direction, although the estimated gap remains smaller. For fare evasion, tax evasion, and bribery, the combined estimates adjust the LLM-only contrasts, but do not fully recover the positive differences found in human data; History-only performs better in preserving their direction. The predicted cross-country gaps in homosexuality and divorce acceptance also remain smaller than those observed. Overall, these results suggest that incorporating historical information can improve the representation of cross-country differences in LLM-generated responses, although the extent of improvement varies across items. This finding is consistent with previous research documenting challenges in representing cross-national opinions using LLMs \citep{Durmus2024} and highlights the potential value of historical behavioral data in addressing these challenges.

\subsection{Correlation Structure and Response Heterogeneity}\label{subsec:correlations}

Researchers often study the relationships among attitudes, rather than each response in isolation. Agreement in means can conceal a failure to preserve which opinions occur together within an individual. This is especially relevant when synthetic responses are used to explore preference relationships or to compare participant pools \citep{snowberg2021testing}. We therefore examine Pearson correlations within the social-value, trust, and ethical-norm blocks: three, six, and twenty-one item pairs, respectively, for thirty pairs in each country.

\begin{figure}[!ht]
    \centering
    \includegraphics[width=1\linewidth]{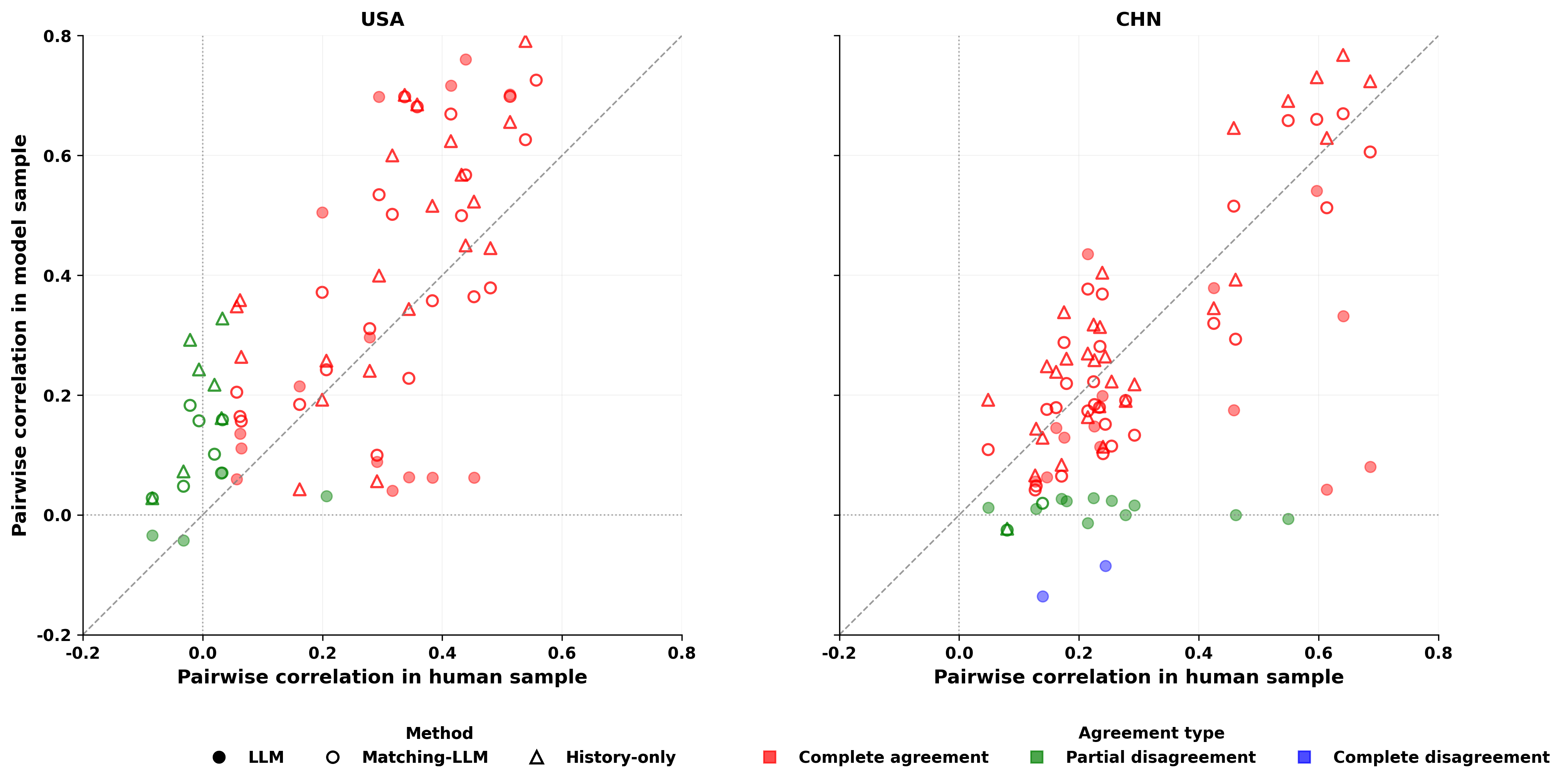}
    \caption{Pairwise correlations in human and model-generated responses.\label{fig:heatmap}}
    \caption*{\footnotesize \setstretch{1} \textit{Notes: }The x-axis reports the human-sample correlation and the y-axis the corresponding model-sample correlation for a pair of items within the social-values, trust, or ethical-norm block. Filled circles denote LLM-only, open circles Matching-LLM, and open triangles History-only. Marker color indicates agreement: red for significant correlations of the same sign in both samples, green for significance in only one sample, and blue for significant correlations of opposite signs. Significance is assessed using two-sided tests at the 10\% level, consistent with the plotted classification. The dashed 45-degree line indicates equality of correlation coefficients; dotted horizontal and vertical lines mark zero. Pairs with undefined correlations, including those involving constant model responses, are omitted separately for each method. The U.S. panel contains 19 LLM-only pairs and 30 pairs for each of History-only and Matching-LLM; the China panel contains 27 LLM-only pairs and 30 pairs for each of the other methods. The human samples contain 2{,}077 U.S. and 2{,}001 Chinese respondents.}
\end{figure}

In Figure~\ref{fig:heatmap}, the horizontal axis gives the human correlation and the vertical axis the corresponding method correlation. Filled circles denote LLM-only, open circles Matching-LLM, and open triangles History-only. Marker colors classify agreement using the supplied two-sided 10\% significance threshold. The classifications indicate whether the methods reproduce the direction and statistical significance of relationships among attitudes, providing a summary of agreement in correlation patterns. Distance from the 45-degree line provides complementary information about coefficient magnitudes.

Furthermore, Figure~\ref{fig:heatmap} shows that incorporating historical information helps preserve the correlation structure observed in human survey responses. Both History-only and Matching-LLM yield computable correlations for all thirty item pairs in each country, compared with nineteen in the United States and twenty-seven in China for LLM-only, where nearly constant responses make some correlations undefined. In the United States, LLM-only achieves complete agreement with human correlation patterns for fifteen of its nineteen computable pairs, while History-only and Matching-LLM each achieve complete agreement for twenty-three of thirty pairs. In China, LLM-only achieves complete agreement for fourteen of twenty-seven pairs, with two pairs exhibiting significant correlations of opposite signs. By comparison, History-only and Matching-LLM achieve complete agreement for twenty-nine and twenty-eight of thirty pairs, respectively, with no significantly opposite-signed correlations. These results demonstrate that incorporating historical responses substantially improves the representation of the correlation patterns observed in human data.

Comparisons of correlation magnitudes provide further insights into the performance of Matching-LLM. In the United States, the average absolute correlation gap decreases from 0.178 for History-only to 0.141 for Matching-LLM, indicating that incorporating LLM-generated information can further improve the accuracy of historical predictions. In China, the corresponding gaps are 0.084 and 0.085, suggesting that the combined approach achieves accuracy comparable to the already strong historical baseline. Overall, these findings highlight the value of historical anchoring in improving the correlation structure of synthetic responses. They also suggest that the additional benefits of combining historical and LLM-generated information, to some extent, depend on the predictive strength of the historical baseline.

\subsection{Cross-Item Transfer Validation}\label{subsec:transfer}

The fitted results cannot establish that the historical weight transfers to other outcomes. To examine that issue, we apply the country-specific weights from Equation~\eqref{eq:wvs-calibration} to six gender-role and family-value questions, Q27--Q32, which were excluded from the fourteen-item calibration set. Matching and the demographic profiles remain the same. The weight is not re-estimated on these six items, so their observed responses provide a test of cross-item transfer, rather than an additional calibration target.

\begin{figure}[!ht]
    \centering
    \includegraphics[width=1\linewidth]{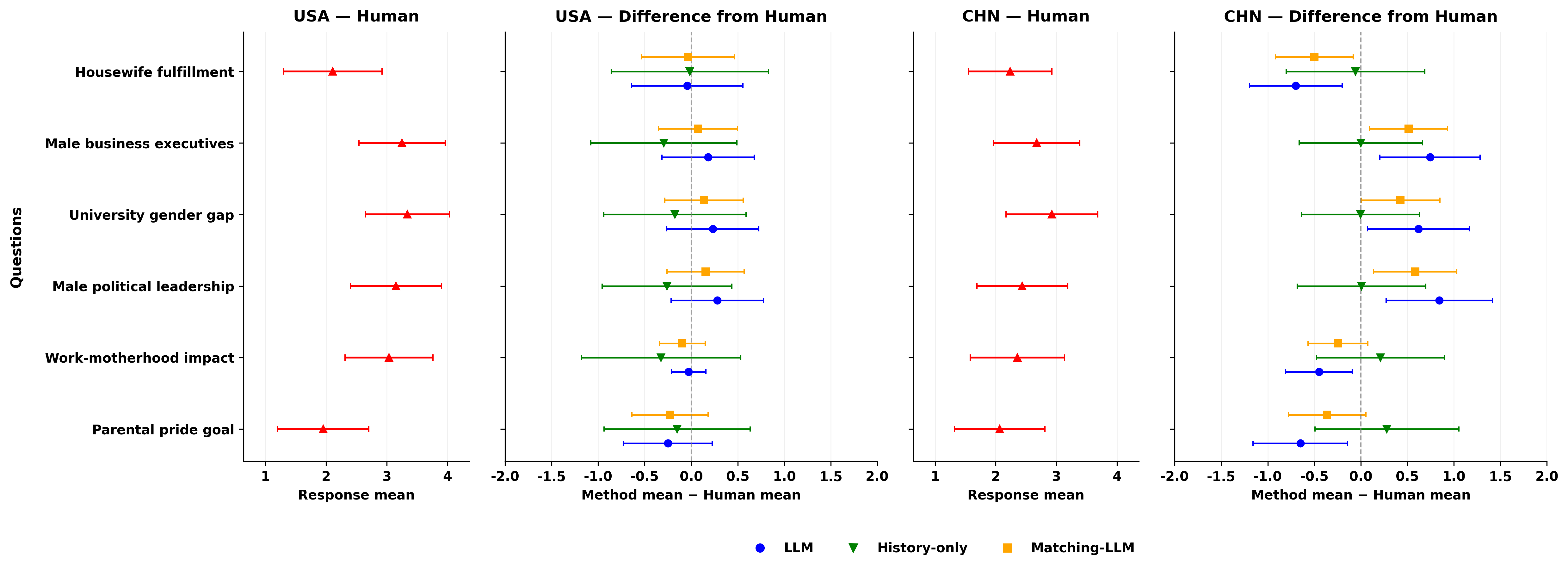}
    \caption{Cross-item validation: human means and method deviations.\label{fig:new-qustions-WVS}}
    \caption*{\footnotesize \setstretch{1} \textit{Notes: }The six items were not used to estimate the historical weights. For each country, the human-benchmark column shows WVS Wave 7 response means as red upward triangles, with horizontal bars extending one standard deviation on either side. The adjacent column reports each method's mean minus the human mean: blue circles denote LLM-only, green downward triangles History-only, and orange squares Matching-LLM. Zero indicates agreement in means; bars in the deviation columns show the respective method's standard deviation. History-only uses matched Wave 6 responses, and Matching-LLM applies the historical weights estimated from the calibration items: 0.23 for the U.S. ($N=2{,}077$) and 0.31 for China ($N=2{,}001$). All six items use 4-point response scales. Full prompts are provided in Supplementary Appendix C.}
\end{figure}

Figure~\ref{fig:new-qustions-WVS} shows that Matching-LLM is closer to the human mean than LLM-only in eleven of the twelve country--item comparisons. The exception is work--motherhood impact in the U.S., where LLM-only is already close to the benchmark and the historical adjustment moves the prediction away. Thus, the gains on the original items extend to many unused outcomes, but not automatically to every one.

The comparison with History-only reveals a difference between the two countries. In the United States, Matching-LLM improves on LLM-only for five of the six unused items and also outperforms History-only on several questions, providing further evidence of the value of combining historical and LLM-generated information. In China, Matching-LLM improves on LLM-only across all six items, although History-only remains closer to the human means. For example, on the male business executives item, the human and historical means are nearly identical at 2.675 and 2.674, compared with 3.419 for LLM-only and 3.188 for Matching-LLM. A similar pattern appears for the university gender-gap item, where the human and historical means are 2.926 and 2.921, respectively. These results suggest that the combined approach can improve predictions beyond the original calibration questions, while its incremental gains over historical data depend on the predictive strength of the historical baseline.

The difference between the U.S. and Chinese results also raises an interesting question about the cross-cultural performance of LLMs. As GPT-4o is developed in the United States, one possibility is that it captures certain patterns in U.S. social attitudes more effectively, allowing its predictions to provide greater incremental value when combined with historical data. Differences in the stability of historical responses and the characteristics of individual questions may also contribute to this pattern. Although the present analysis does not distinguish among these explanations, the results suggest that the benefits of combining historical and LLM-generated information may vary across countries. Finally, because this validation uses questions from the same survey wave, it primarily evaluates prediction across different survey items. We next extend the analysis to a different setting by applying historically calibrated information to subsequent election outcomes, providing a further assessment of Matching-LLM in forecasting over time.

\section{Election Forecasting Applications}\label{sec:elections}\label{subsec:prediction2}

Election outcomes provide externally observed targets beyond the survey questions used for WVS calibration. We first examine historical-weight calibration against the 2020 U.S. election, then compare the three methods in the 2024 application, and finally report the German application as supplementary evidence across electoral systems. The U.S. analysis compares party vote-share deviations, statewide winning parties, and national electoral-vote totals. The German exercise uses party-specific multiplicative calibration and compares the forecasts with the observed parliamentary distribution and state-leading parties.

The U.S. simulations were completed on or before October 30, 2024, and the German simulations on or before February 10, 2025. Public archive dates provide a separate record of disclosure: arXiv version 1 was submitted on November 3, 2024, before the November 5 U.S. election,\footnote{See \url{https://arxiv.org/abs/2411.01582v1} for the first version of the draft.} and version 2 on February 11, 2025, before the February 23 German election.\footnote{See \url{https://arxiv.org/abs/2411.01582v2} for the second version of the draft. The simulation-completion dates and public submission dates are distinct.} These public records document the disclosure of forecasts before the election outcomes were known. \footnote{The pre-election draft with arXiv version 1 used a historical weight of $h=0.80$, whereas the specification reported here uses $h=0.82$ from the MSE-based 2020 calibration. The two weights imply the same statewide winning parties and national electoral-vote totals: 310 for the Republican candidate and 228 for the Democratic candidate.}

\subsection{Historical Weight Calibration Using the 2020 U.S. Election}\label{subsec:2020-calibration}

\begin{figure}[!htbp]
    \centering
    \includegraphics[width=\linewidth]{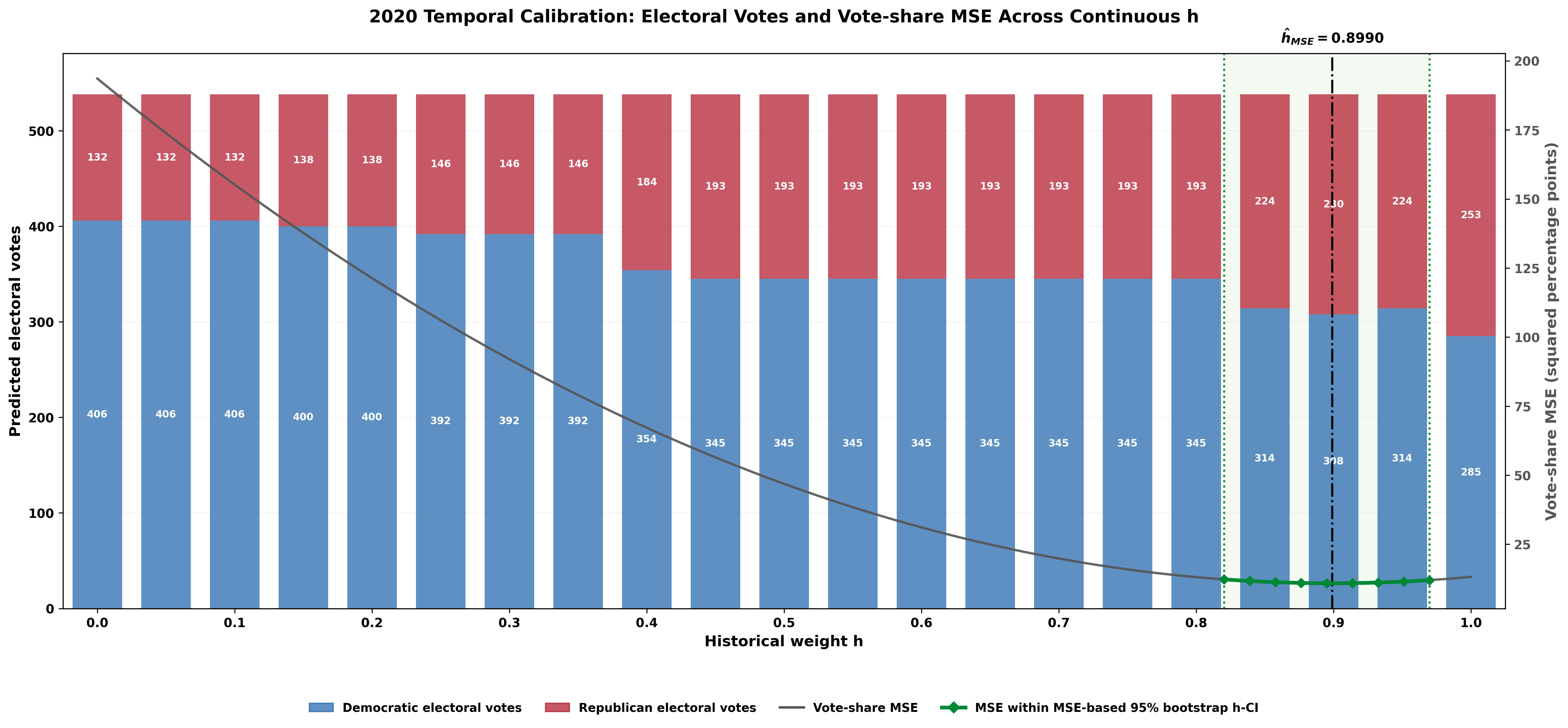}
    \caption{2020 historical-weight calibration and sensitivity analysis.\label{fig:h_sensitivity}}
    \caption*{\footnotesize \setstretch{1} \textit{Notes: }The horizontal axis is the historical weight $h\in[0,1]$. The historical baseline averages the 2012 and 2016 state-level party vote shares; actual 2020 outcomes serve as calibration targets. Blue and red stacked bars show predicted Democratic and Republican electoral votes at selected weights (left axis), summing to 538. The gray line reports continuous vote-share MSE in squared percentage points (right axis). The black dash-dotted line marks the MSE-minimizing weight, $\widehat h_{\mathrm{MSE}}=0.8990$. Green dotted lines and shading mark the 95\% percentile bootstrap interval, approximately $[0.8203,0.9698]$; green diamonds highlight the MSE curve within this weight interval. The interval is based on 10{,}000 jurisdiction-level resamples and summarizes sensitivity to calibration-sample composition, conditional on the available inputs. The 2024 application uses $h=0.82$, the lower endpoint rounded to two decimal places.}
\end{figure}

We minimize vote-share MSE over the continuous interval $h\in[0,1]$, using the historical average of 2012 and 2016 state-level party vote shares, simulated 2020 LLM shares, and actual 2020 outcomes. This sequence prevents the historical baseline from including its own calibration target. The loss in Equation~\eqref{eq:election-mse} averages squared errors over Democratic and Republican shares in the fifty states and the District of Columbia. Figure~\ref{fig:h_sensitivity} presents the continuous MSE curve across the full range of historical weights, together with predicted electoral-vote totals at selected values of $h$.

Figure~\ref{fig:h_sensitivity} shows a substantial decline in vote-share MSE as historical information is added. MSE falls from 193.594 squared percentage points at $h=0$ to its continuous minimum of 10.900 at $\widehat h_{\mathrm{MSE}}=0.8990$, before rising to 13.206 at $h=1$. The calibrated combination therefore improves on both single-source predictors under the calibration criterion, although its improvement over History-only is considerably smaller than its improvement over LLM-only.

The bootstrap distribution from 10{,}000 jurisdiction-level resamples gives a 95\% interval of approximately $[0.8203,0.9698]$. We use its lower endpoint rounded to two decimal places, $h=0.82$, for the 2024 application. Evaluated against the 2020 calibration target, this application weight gives an MSE of 12.310 squared percentage points.

 At selected $h=0.82$, the 2020 calibration produces 345 Democratic and 193 Republican electoral votes, an absolute error of thirty-nine relative to the actual Republican total of 232, and correctly classifies forty-seven of fifty-one jurisdictions (92.16\%).

\subsection{Forecasting the 2024 U.S. Presidential Election}\label{subsec:2024-forecast}

\begin{figure}[!htbp]
    \centering
    \begin{subfigure}[t]{0.48\linewidth}
        \centering
        \includegraphics[width=\linewidth]{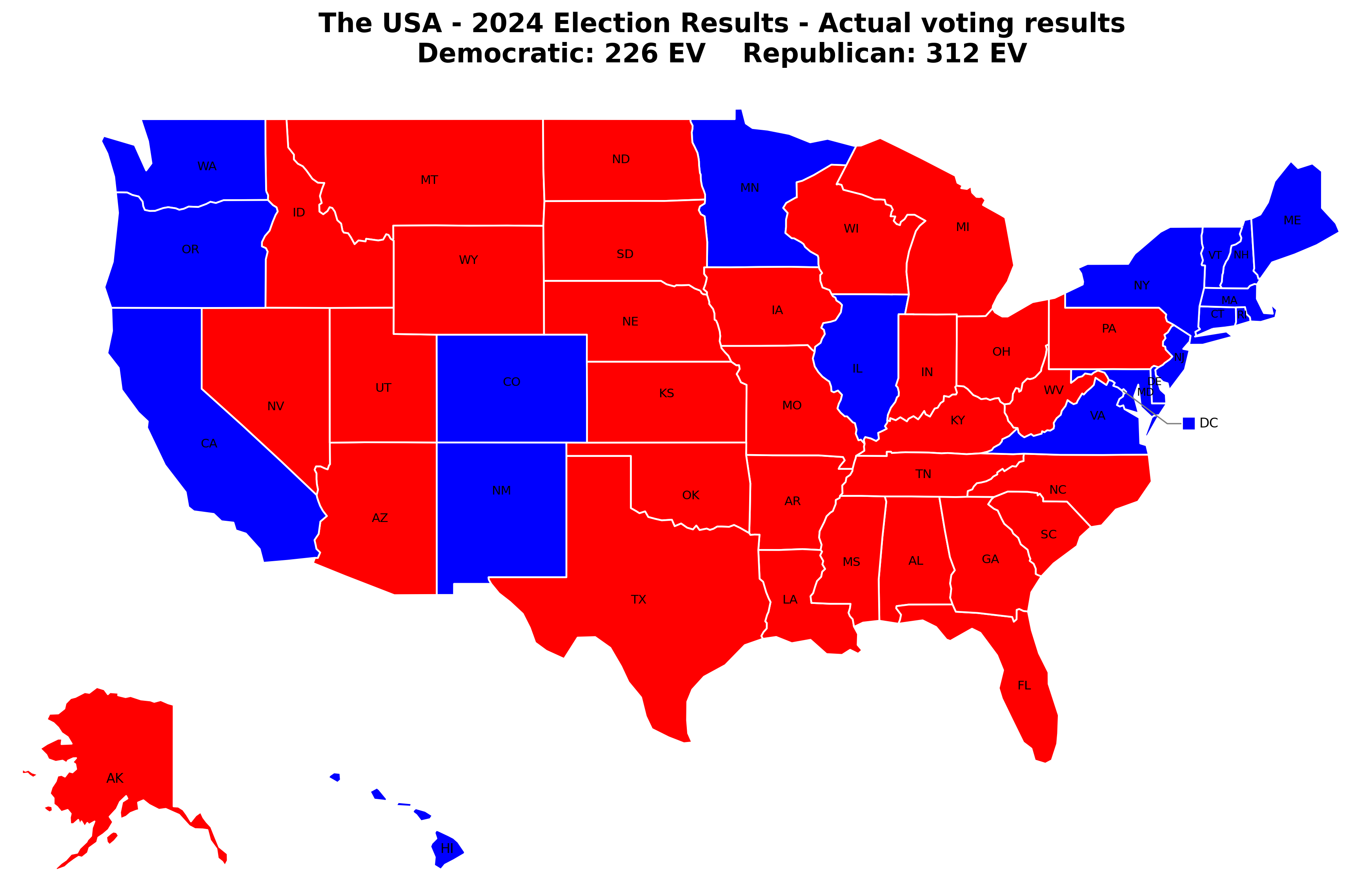}
        \caption{Actual voting results}
    \end{subfigure}
    \hfill
    \begin{subfigure}[t]{0.48\linewidth}
        \centering
        \includegraphics[width=\linewidth]{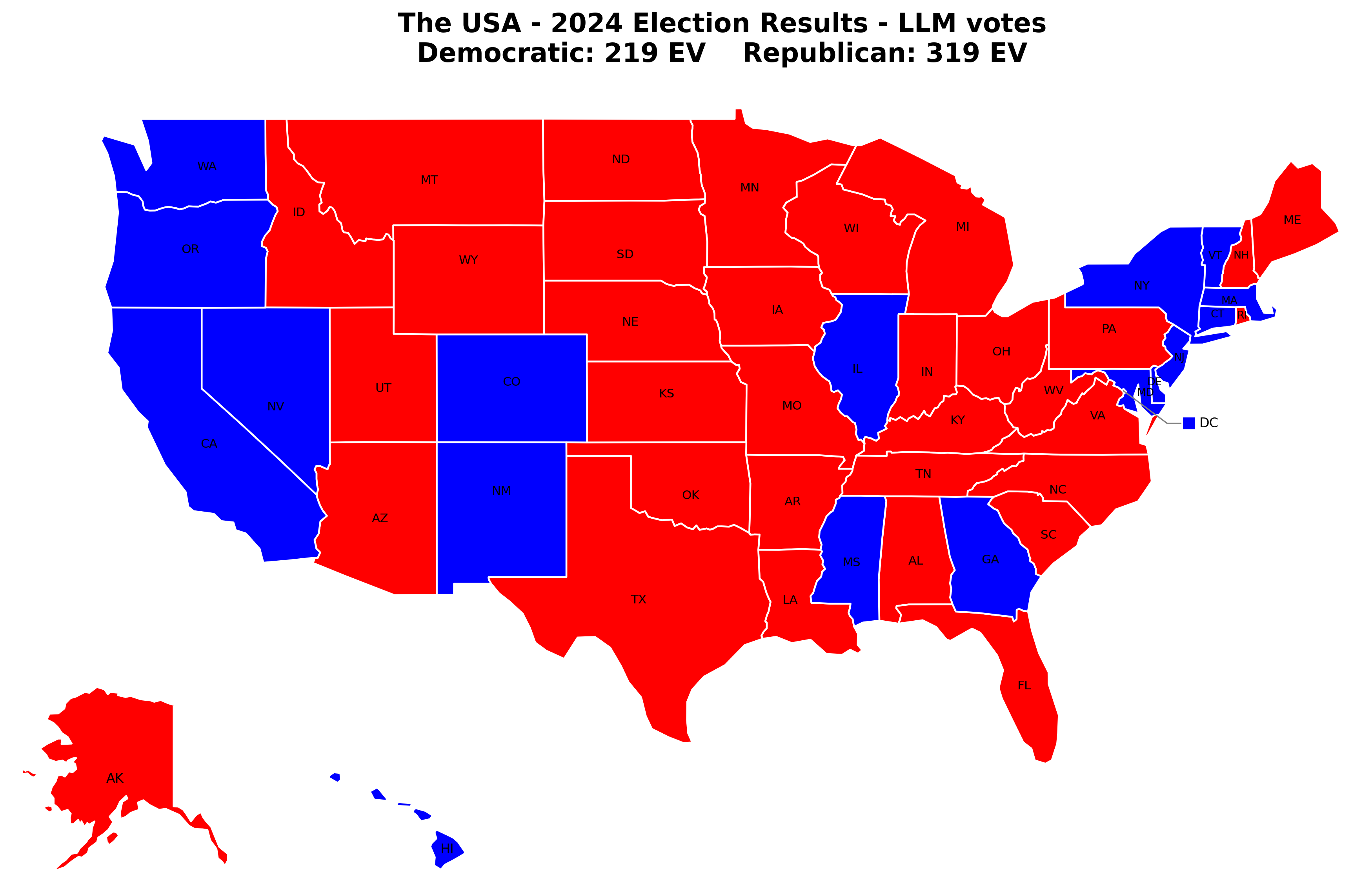}
        \caption{LLM-only}
    \end{subfigure}
    \par\medskip
    \begin{subfigure}[t]{0.48\linewidth}
        \centering
        \includegraphics[width=\linewidth]{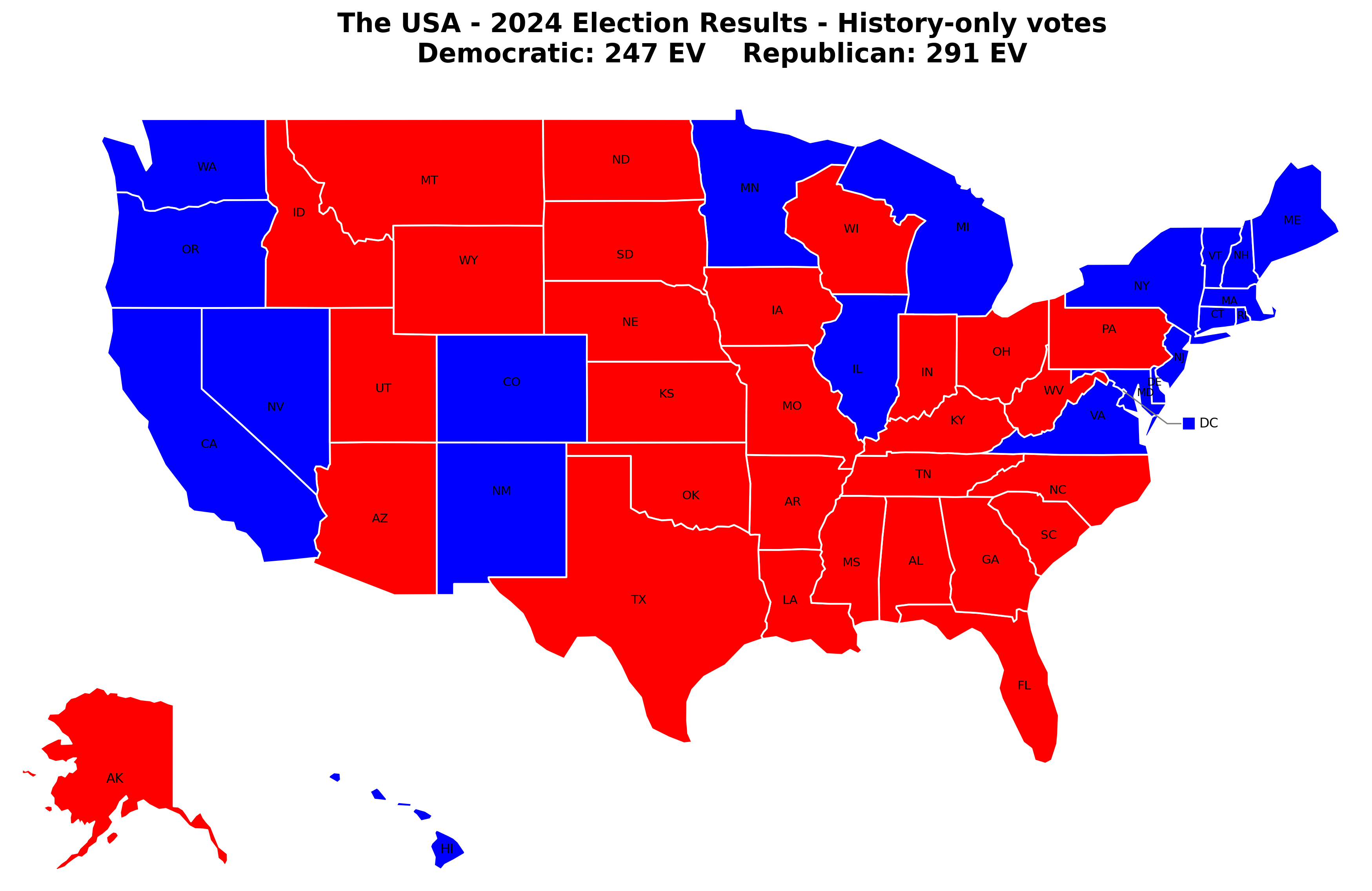}
        \caption{History-only}
    \end{subfigure}
    \hfill
    \begin{subfigure}[t]{0.48\linewidth}
        \centering
        \includegraphics[width=\linewidth]{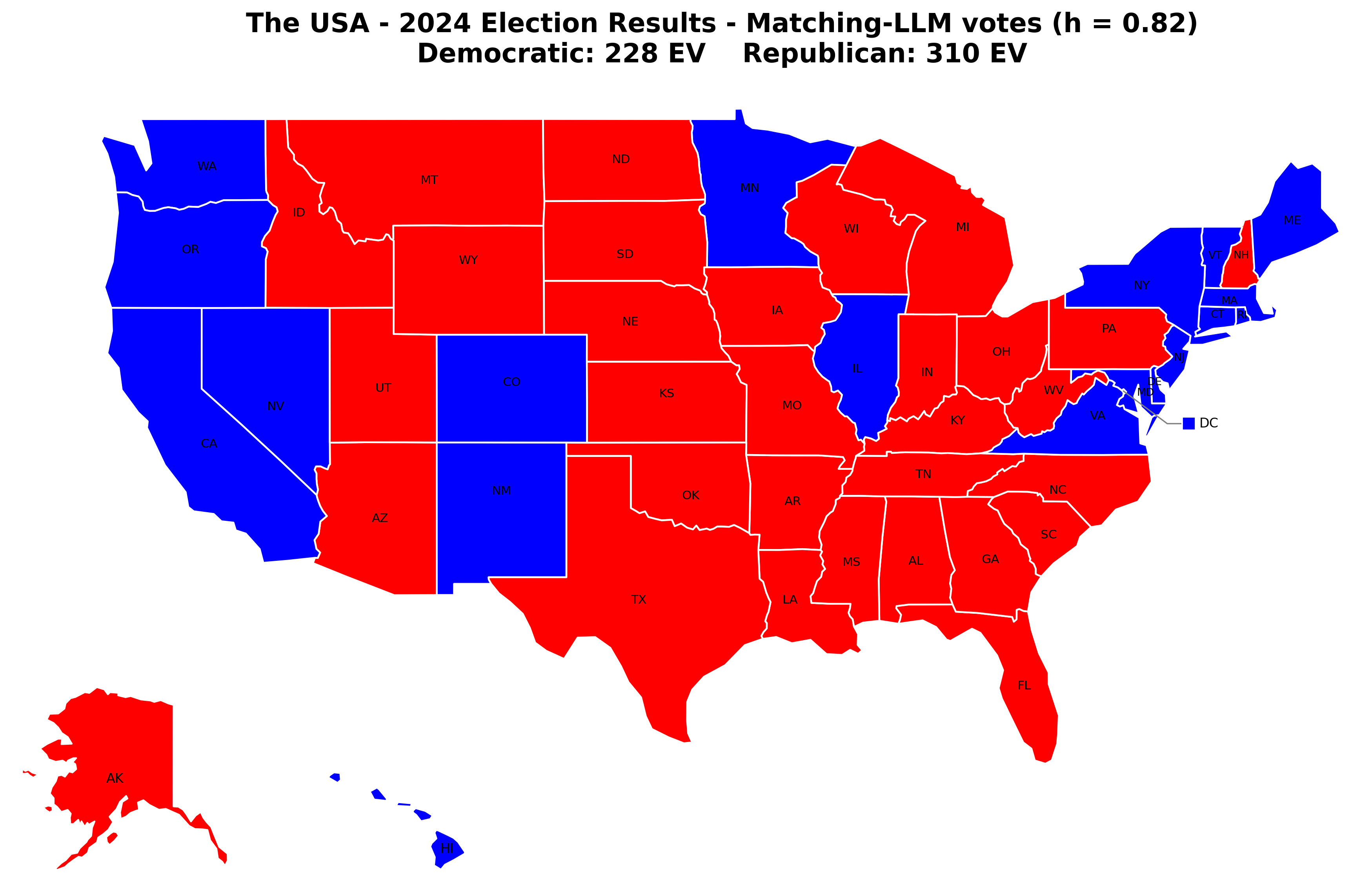}
        \caption{Matching-LLM ($h=0.82$)}
    \end{subfigure}
    \caption{Actual and predicted outcomes of the 2024 U.S. presidential election.\label{fig:2024_election}}
    \caption*{\footnotesize \setstretch{1} \textit{Notes: }Blue states indicate Democratic statewide winners and red states Republican statewide winners. (a) Actual result: 226 Democratic and 312 Republican electoral votes. (b) LLM-only: 219 Democratic and 319 Republican electoral votes, based on simulated votes from the ANES 2020 demographic profiles ($N=6{,}571$); the eight misclassified states are Georgia, Maine, Minnesota, Mississippi, Nevada, New Hampshire, Rhode Island, and Virginia. (c) History-only: 247 Democratic and 291 Republican electoral votes, using the average of actual 2016 and 2020 state-level party vote shares; Michigan and Nevada are misclassified. (d) Matching-LLM: 228 Democratic and 310 Republican electoral votes, combining the baseline in (c) with the LLM-only vote shares in (b) at weights of 0.82 and 0.18; Nevada and New Hampshire are misclassified. The actual results account for congressional-district electoral-vote allocations in Maine and Nebraska, whereas all three forecasting methods use statewide winners without separately modeling congressional-district outcomes.}
\end{figure}

For the 2024 forecast, LLM-only votes are generated using the 6{,}571 ANES 2020 profiles and the Trump--Harris candidate pairing. History-only uses the average of actual 2016 and 2020 state-level party shares. Matching-LLM assigns weights of 0.82 and 0.18 to the historical and LLM signals, respectively. Figure~\ref{fig:2024_election} displays actual voting, LLM-only, History-only, and Matching-LLM. The actual Electoral College outcome is 312 Republican and 226 Democratic votes \citep{NARA2024}.

All three methods identify Trump as the national winner, so winner correctness alone does not discriminate between them. LLM-only predicts 319 Republican electoral votes, History-only 291, and Matching-LLM 310. The corresponding absolute errors are seven, twenty-one, and two. The combination therefore improves this aggregate metric relative to both endpoints with the selected weight.

State-level classification provides a different perspective. LLM-only misclassifies eight states: Georgia, Maine, Minnesota, Mississippi, Nevada, New Hampshire, Rhode Island, and Virginia. History-only misclassifies Michigan and Nevada; Matching-LLM misclassifies Nevada and New Hampshire. Both history-based methods thus correctly classify forty-nine of fifty-one jurisdictions. Adding the LLM signal changes Michigan to the correct Republican winner but changes New Hampshire to an incorrect Republican winner; Nevada remains assigned to the Democrats. This comparison establishes an improvement in the allocation of electoral votes without a reduction in the number of state errors relative to History-only.

Among the seven commonly identified swing states---Arizona, Georgia, Michigan, Nevada, North Carolina, Pennsylvania, and Wisconsin---History-only misses Michigan and Nevada, while Matching-LLM misses Nevada alone. The LLM contribution is consequently useful for the composition of correctly classified swing states, even though it does not improve total state-winner accuracy. 

The results also highlight the importance of considering different measures of electoral prediction accuracy. Although LLM-only misclassifies eight states, its electoral-vote error is relatively small because errors across states can offset one another in the national total. Conversely, misclassifying fewer states does not necessarily lead to a smaller electoral-vote error, as states differ in their electoral-vote allocations. The maps report statewide winners, and our predictions for Maine and Nebraska are also based on statewide outcomes without separately modeling congressional-district results. These differences illustrate why state-level accuracy and aggregate electoral-vote accuracy provide complementary assessments of forecasting performance.

Overall, these results suggest that incorporating LLM-generated information can improve certain aspects of election forecasting relative to the historical baseline. However, the analysis does not identify whether candidate-specific factors, issue salience, or other contextual changes account for the differences in state-level predictions.

\subsection{Forecasting the 2025 German Federal Election}\label{subsec:germany}

The German application is a broader extension of Matching-LLM in which historical information calibrates LLM-generated predictions through party-specific adjustments. We present it as a supplementary application in a multiparty electoral setting, comparing the forecasts with observed election outcomes. The forecasts were completed before the February 23, 2025 German election. The simulations use 3{,}965 demographic and political profiles from the 2021 German Longitudinal Election Study (GLES) pre-election cross-sectional survey, which provides detailed voter information \citep{GLES2023}. Party-specific multiplicative calibration coefficients are obtained by comparing LLM simulations with historical outcomes from 2017 and 2021, then applied to the 2025 simulations. These adjustments are distinct from the common convex historical weight used in the U.S. application; Germany is therefore an extension of historical calibration, rather than a replication of the convex-weight specification used in the U.S. presidential application.

\begin{figure}[htbp]
  \centering
  \begin{subfigure}{\textwidth}
    \centering
    \includegraphics[width=0.78\textwidth,height=0.34\textheight,keepaspectratio]{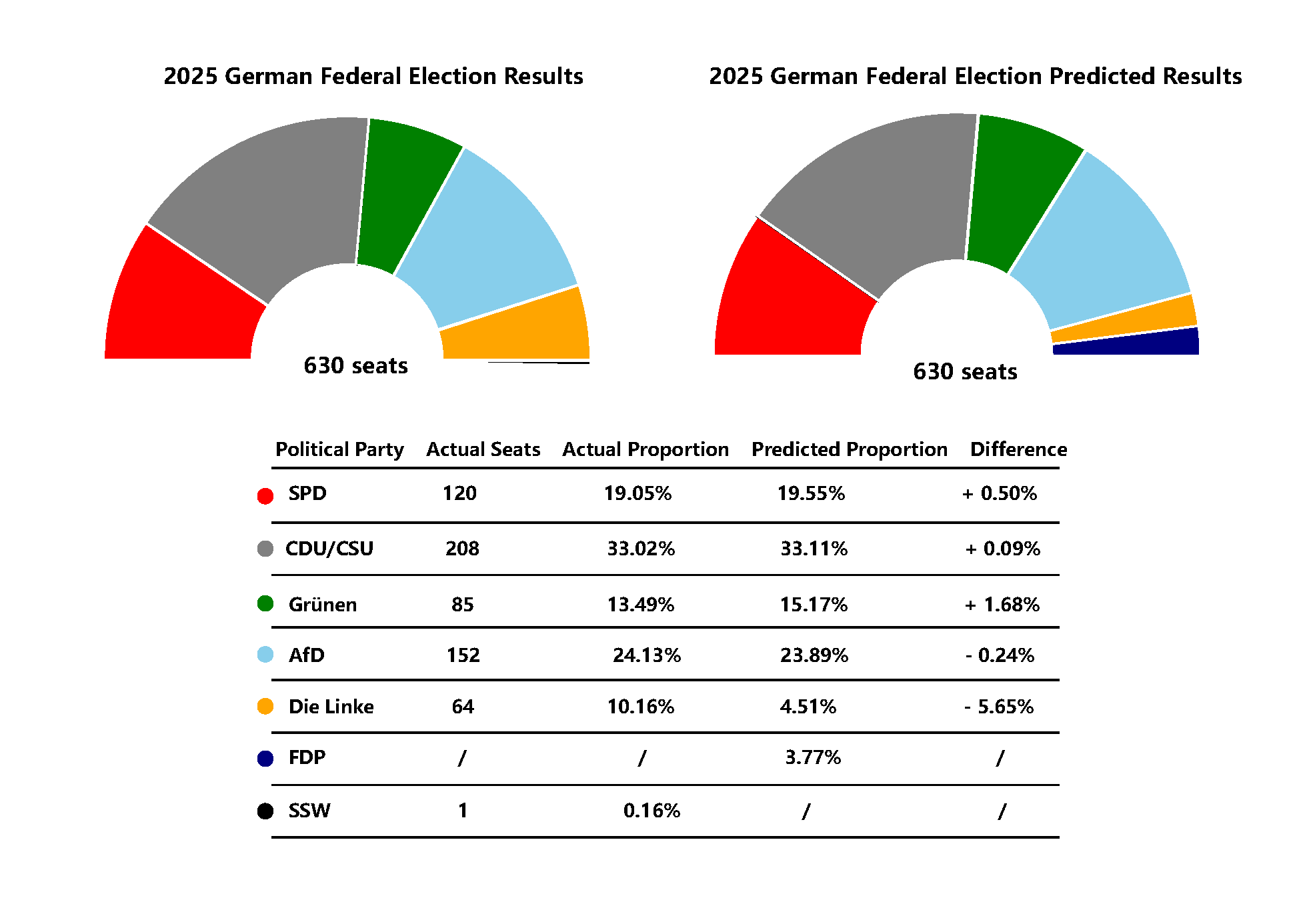} 
    \caption{Distribution of party seats in the Bundestag}
    \label{fig:top}
  \end{subfigure}
  \begin{subfigure}{0.8\textwidth}
      \begin{subfigure}{0.4\textwidth}
        \centering
        \includegraphics[width=\textwidth,height=0.25\textheight,keepaspectratio]{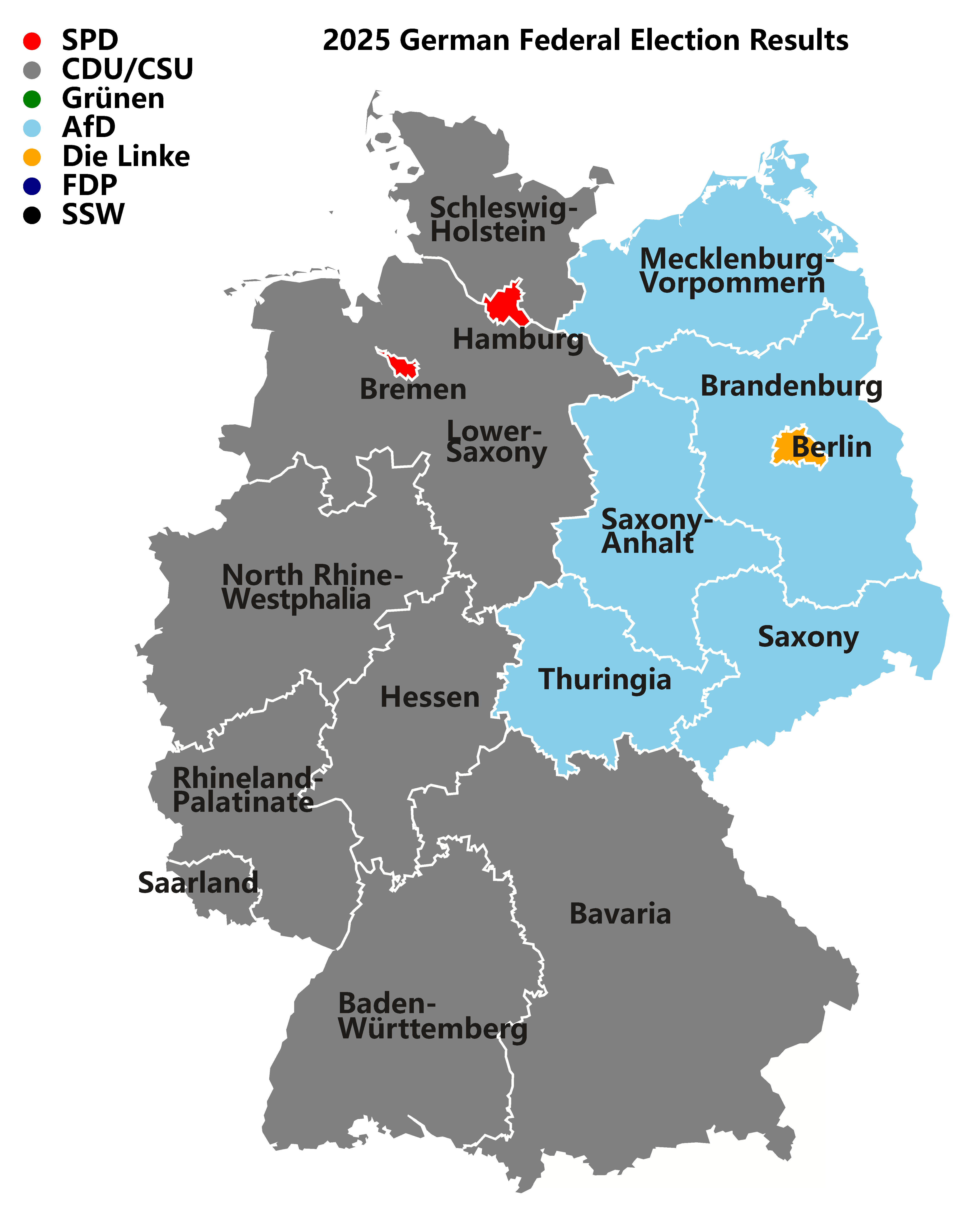} 
        \label{fig:bottom_left}
      \end{subfigure}
      \hfill
      \begin{subfigure}{0.4\textwidth}
        \centering
        \includegraphics[width=\textwidth,height=0.25\textheight,keepaspectratio]{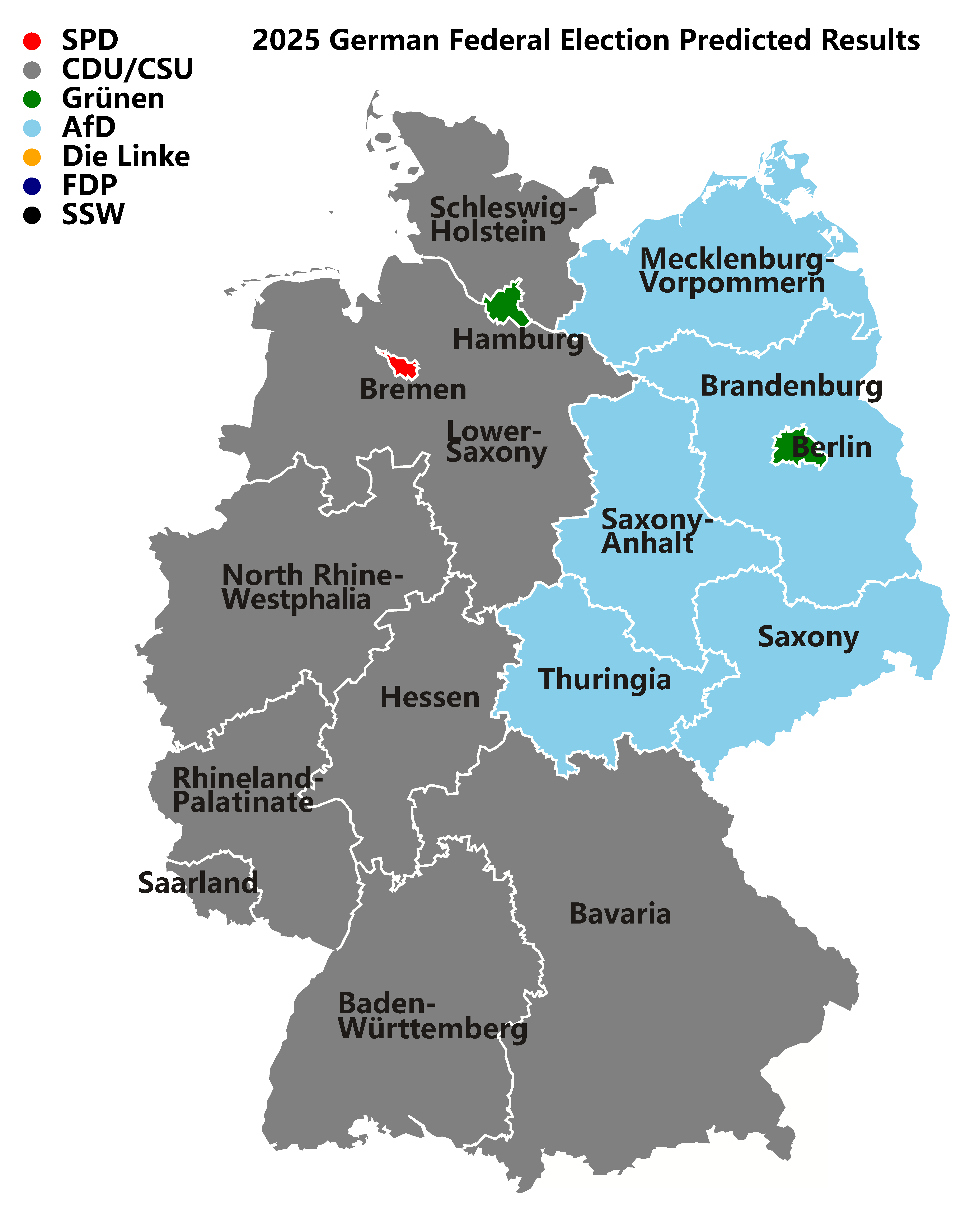} 
        \label{fig:bottom_right}
      \end{subfigure}
   \caption{Highest-polling party in each state}
  \end{subfigure}
  \caption{Prediction of the 2025 German Federal Election\label{fig:German}}
  \caption*{\footnotesize \setstretch{1} \textit{Notes:} The figure compares Matching-LLM forecasts with observed outcomes for the 2025 German federal election, using GLES profiles ($N=3{,}965$). The simulations model the second vote for six major parties, with party-specific multiplicative calibration coefficients obtained from 2017 and 2021 outcomes. The upper panel presents the predicted party proportions alongside the actual Bundestag seat distribution. The lower panels compare the highest-polling party in each federal state; fourteen of sixteen are predicted correctly, with Hamburg and Berlin mispredicted.}
\end{figure}

After applying the party-specific historical adjustments, the Matching-LLM forecasts in Figure~\ref{fig:German} project approximately 19.55\% for SPD, compared with its actual parliamentary representation of 120 seats (19.05\% of the Bundestag). CDU/CSU is predicted to lead with about 33.11\%, compared with 208 seats (33.02\%). The Greens are forecast at 15.17\%, compared with 85 seats (13.49\%), and FDP at 3.77\%, although it receives no Bundestag seats. AfD is predicted to receive 23.89\%, compared with 152 seats (24.13\%), and Die Linke 4.51\%, compared with 64 seats (10.16\%).\footnote{In our analysis, we compared the predicted vote shares to the actual parliamentary seat distribution because our predictive model simplified and normalized the voting process, focusing solely on the six major political parties and excluding smaller parties that did not enter the Bundestag.} The forecast identifies CDU/CSU, AfD, and SPD as the three leading parties in the correct order, with the largest discrepancy among the represented parties occurring for Die Linke.

At the state level, Matching-LLM correctly identifies the party with the highest vote share in fourteen of sixteen federal states. In Hamburg, SPD receives the most votes, while the forecast places it second. In Berlin, Die Linke wins the plurality, while the forecast places it third. In both states, the Greens are incorrectly predicted to be the largest party. Although the Greens receive relatively strong support in Hamburg and Berlin compared with other federal states, this does not imply that their vote share exceeds that of SPD in Hamburg or Die Linke in Berlin.

\subsection{Forecasting the 2026 U.S. Senate Election}\label{subsec:2026-senate}

Whereas the German application extends historical calibration to a multiparty setting, the 2026 U.S. Senate application extends Matching-LLM beyond GPT-4o to two additional models, DeepSeek and GPT6-luna. We use 31{,}341 respondent profiles from the 2024 Cooperative Election Study (CES), covering the 33 states with regularly scheduled Senate elections. Each model simulates a choice between the two candidates presented for the respondent's state, conditional on thirteen demographic, socioeconomic, and geographic characteristics. Age is calculated for 2026, and candidate order is randomized across respondents. The profile variables and prompts are provided in Supplementary Appendices A.4 and D.

Simulated choices are aggregated within each state using the CES survey weight to obtain the LLM vote shares. For each model, Matching-LLM combines these shares with the corresponding historical Senate vote shares using the historical weight $h=0.82$ selected in Section~\ref{subsec:2020-calibration}. Given the six-year interval between regular elections for the same Senate seat, the historical anchor uses only the preceding 2020 election, rather than averaging multiple earlier cycles. The combined forecast therefore assigns 82\% weight to the 2020 historical shares and 18\% to the survey-weighted LLM shares. The candidate with the larger predicted share is assigned the state's contested seat.

The special elections in Florida and Ohio are outside the 33-state simulation. For the national seat totals, both are assigned to the Republican Party under an assumption of continued Republican advantage in these states. The 33 modeled seats are then combined with these two seats and the 65 seats not subject to election in 2026. Independent senators are grouped with their caucus or governing alliance when reporting party-bloc totals.

\begin{figure}[!htbp]
    \centering
    \includegraphics[width=0.98\linewidth]{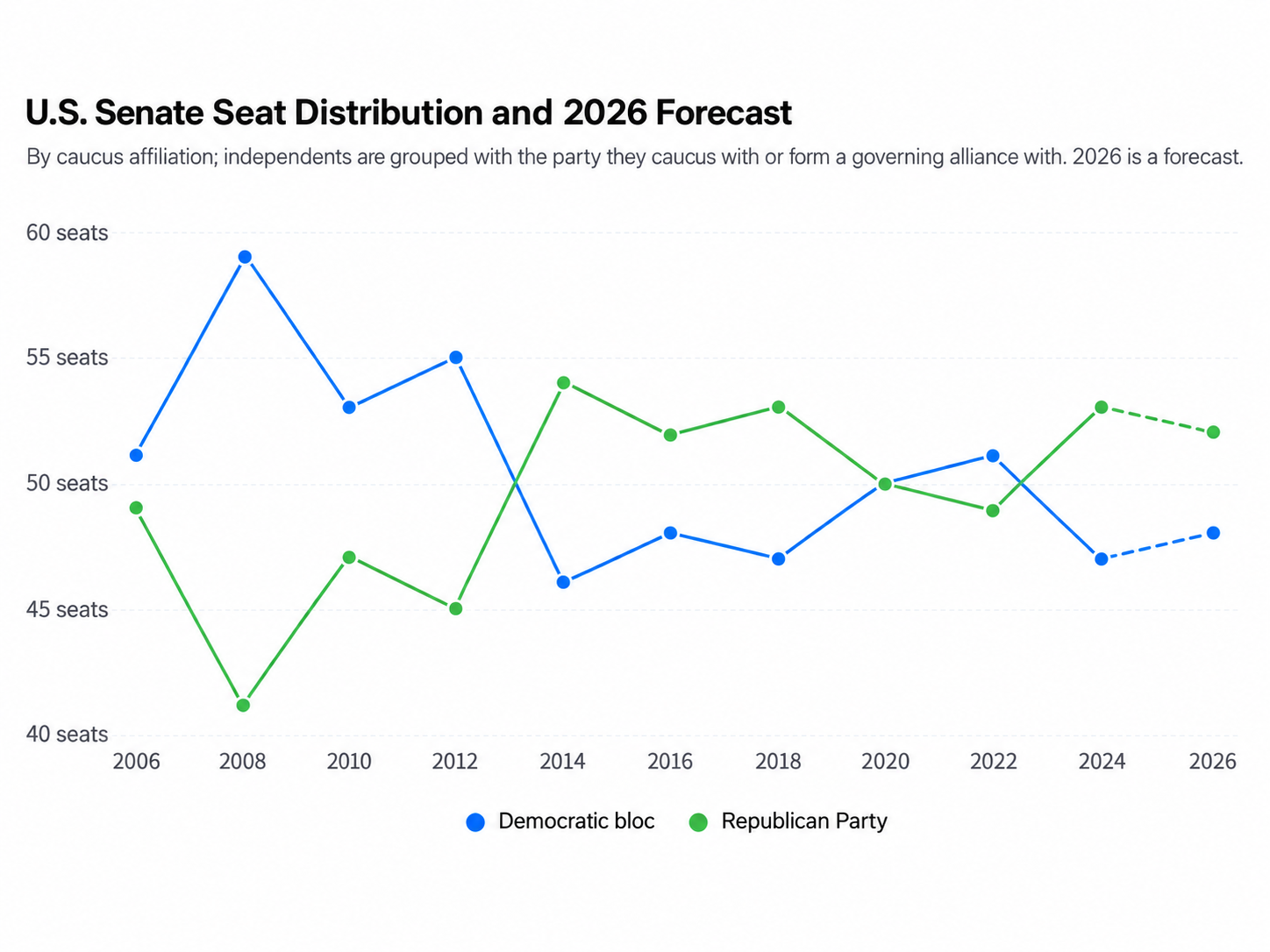}
    \caption{U.S. Senate seat distribution and the 2026 Matching-LLM forecast.\label{fig:2026-senate-trend}}
    \caption*{\footnotesize \setstretch{1} \textit{Notes: }The figure presents party-bloc seat totals from 2006 to 2024 and the common 2026 Matching-LLM forecast from DeepSeek and GPT6-luna: 48 seats for the Democratic bloc and 52 for the Republican Party. Independent senators are grouped with their caucus or governing alliance. Dashed segments connect the 2024 distribution to the 2026 forecast. National forecasts combine the 33 modeled races with the 65 seats not subject to election and the two special-election seats assumed to remain Republican.}
\end{figure}

\begin{figure}[!htbp]
    \centering
    \begin{subfigure}[t]{0.48\linewidth}
        \centering
        \includegraphics[width=\linewidth]{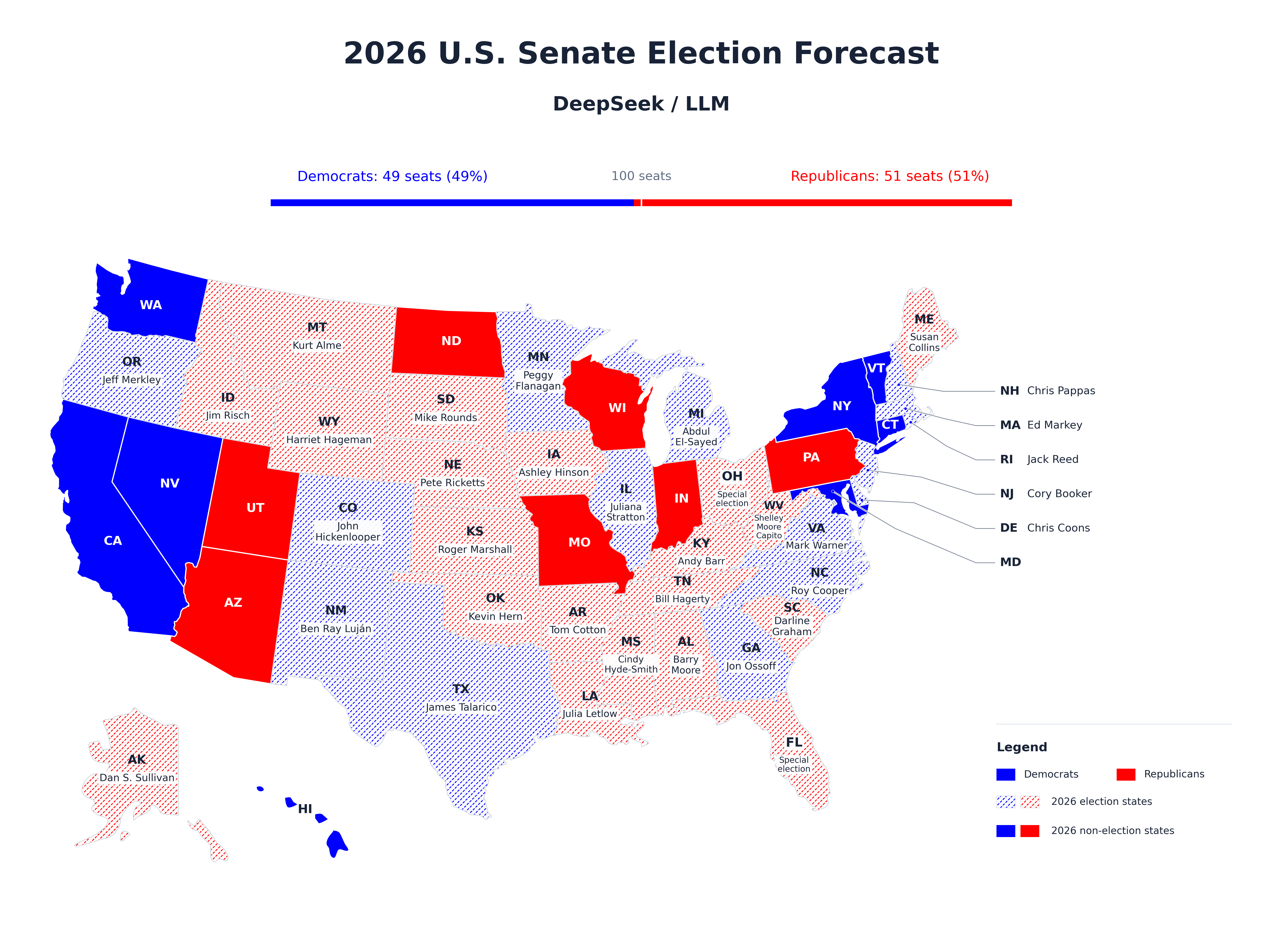}
        \caption{DeepSeek: LLM}
        \label{fig:2026-senate-deepseek-llm}
    \end{subfigure}
    \hfill
    \begin{subfigure}[t]{0.48\linewidth}
        \centering
        \includegraphics[width=\linewidth]{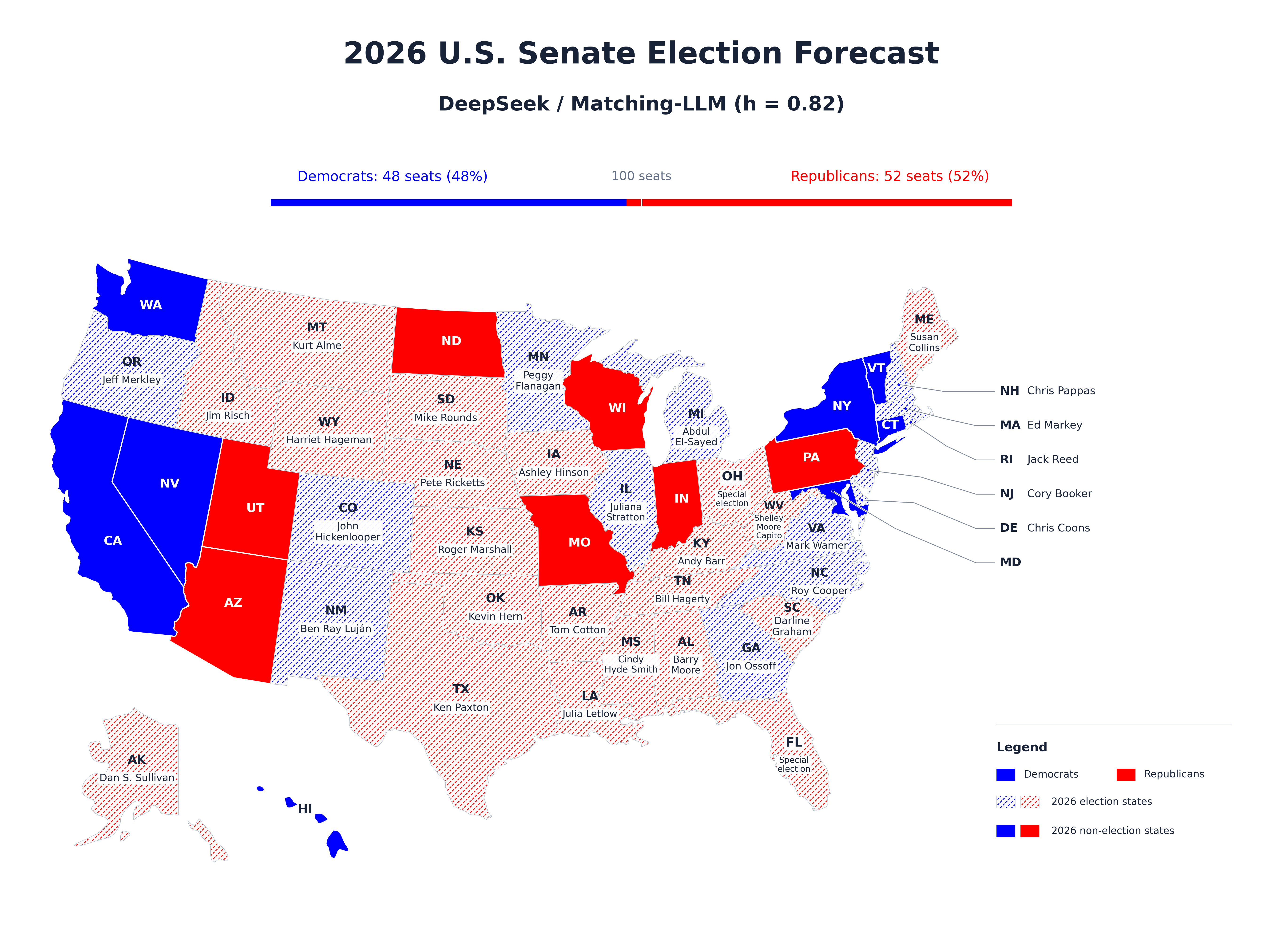}
        \caption{DeepSeek: Matching-LLM}
        \label{fig:2026-senate-deepseek-matching}
    \end{subfigure}
    \par\medskip
    \begin{subfigure}[t]{0.48\linewidth}
        \centering
        \includegraphics[width=\linewidth]{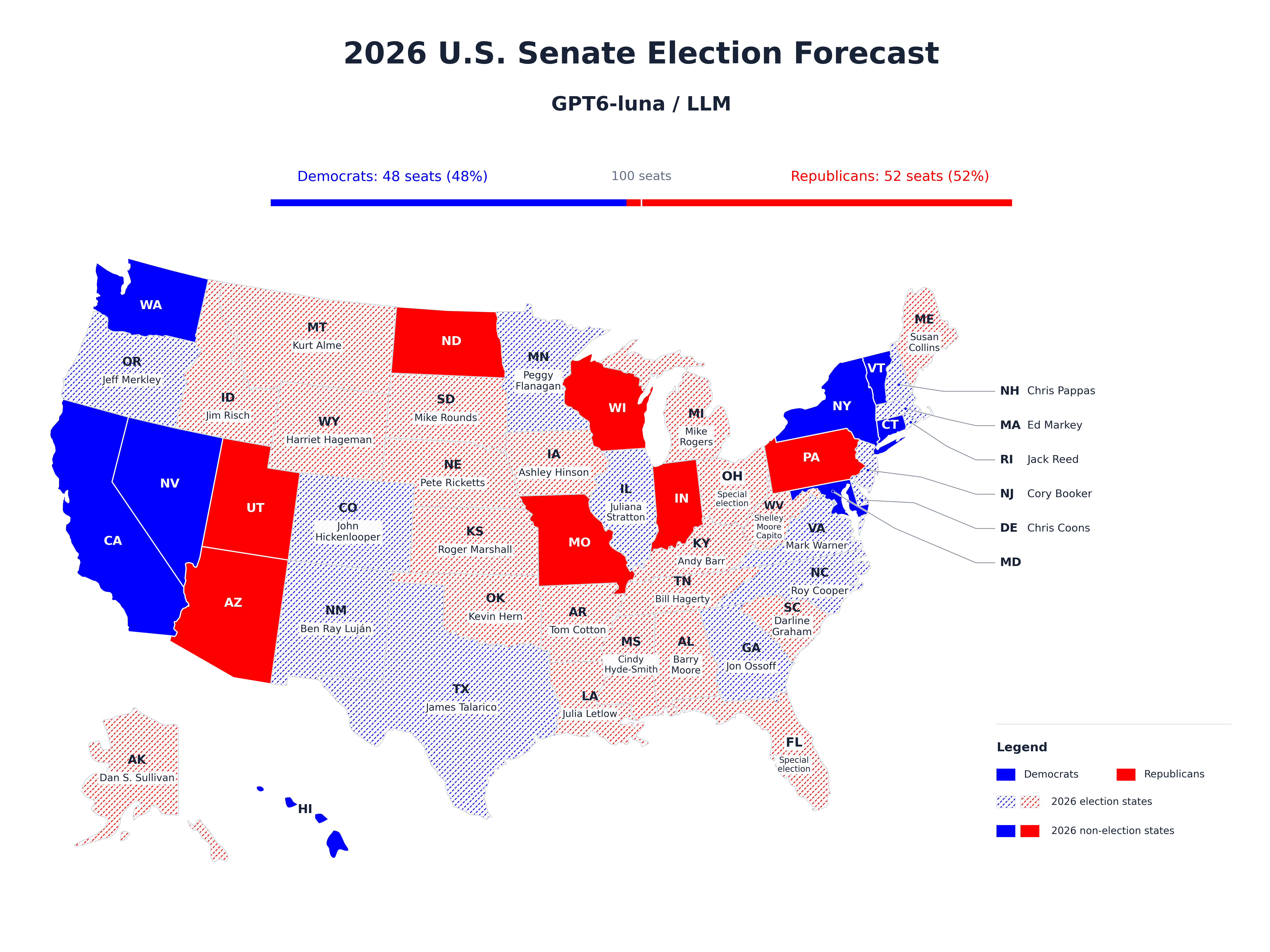}
        \caption{GPT6-luna: LLM}
        \label{fig:2026-senate-gpt6-llm}
    \end{subfigure}
    \hfill
    \begin{subfigure}[t]{0.48\linewidth}
        \centering
        \includegraphics[width=\linewidth]{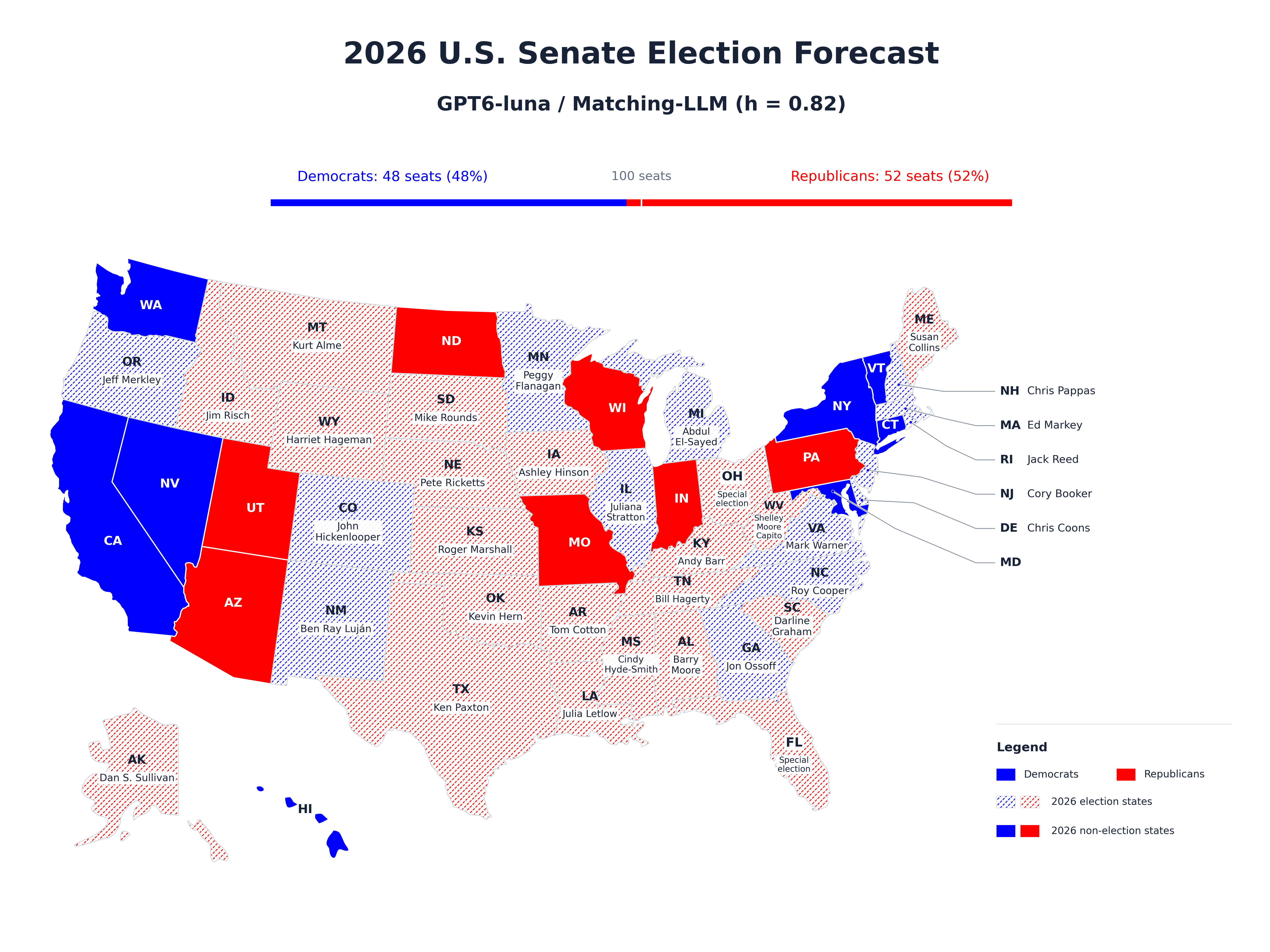}
        \caption{GPT6-luna: Matching-LLM}
        \label{fig:2026-senate-gpt6-matching}
    \end{subfigure}
    \caption{State-level forecasts of the 2026 U.S. Senate election.\label{fig:2026-senate-maps}}
    \caption*{\footnotesize \setstretch{1} \textit{Notes: }Blue and red indicate Democratic and Republican winners, respectively. Hatching identifies states with a 2026 Senate election; candidate names identify the predicted winners in the 33 modeled races. LLM denotes CES survey-weighted simulated vote shares ($N=31{,}341$). Matching-LLM combines these shares with the 2020 Senate shares at a historical weight of $h=0.82$. Panels (a)--(d) imply national Republican--Democratic-bloc seat totals of 51--49, 52--48, 52--48, and 52--48, respectively. These totals include the 65 seats not subject to election and assumed Republican victories in the Florida and Ohio special elections.}
\end{figure}

Figure~\ref{fig:2026-senate-trend} places the common Matching-LLM forecast in the context of the Senate seat distribution since 2006. Both models project 52 Republican seats and 48 seats for the Democratic bloc, compared with 53 and 47, respectively, in the figure's 2024 distribution. Figure~\ref{fig:2026-senate-maps} presents the state-level forecasts for both models and methods. The DeepSeek LLM forecast produces a 51--49 Republican majority, while its Matching-LLM forecast produces 52--48. GPT6-luna produces a 52--48 Republican majority under both methods.

The state-level maps show how similar national totals can arise from different forecasts. In Texas, both LLM forecasts favor the Democratic candidate, whereas both Matching-LLM forecasts favor the Republican candidate. In Michigan, DeepSeek favors the Democratic candidate under both methods, while GPT6-luna changes from a Republican LLM forecast to a Democratic Matching-LLM forecast. For GPT6-luna, these two changes offset in the national total, and, overall, all four forecasts believe that Republicans will have an advantage in the total number of seats.

\clearpage

\section{Discussion}\label{sec:discussion}

\subsection{Complementarity Between Historical and LLM-Based Predictions}

The findings support a historically anchored approach to synthetic prediction while clarifying what its improvement means. Historical records supply observations of realized behavior and its heterogeneity. LLM-generated responses supply predictions conditioned on demographic profiles and the task description. Their combination often improves on an unadjusted LLM, but the relevant empirical question is also whether the generated signal adds value to history. Improvement over LLM-only shows the contribution of historical anchoring; improvement over History-only indicates the additional predictive value of LLM-generated information.

The survey evidence illustrates both outcomes. In the U.S., combining the sources reduces the average gap between synthetic and human correlations relative to History-only. On some mean comparisons, opposing errors also bring the combination close to the benchmark. In China, History-only is already highly informative for the correlation structure and all six unused gender-role and family-value items; the combination does not improve those comparisons. These differences are consistent with the conditional forecast-combination account: improvement depends on the joint error structure, not simply on the fact that sources differ \citep{BatesGranger1969,WangHyndman2023}.

The U.S. election results provide another qualified example. The high historical weight makes the prior-election baseline central to the 2024 prediction. Nevertheless, adding the LLM signal improves the aggregate electoral-vote discrepancy and corrects one important swing-state error. It also introduces an error in New Hampshire, leaving total state accuracy unchanged. Context-conditioned information can therefore improve one decision-relevant aggregate while failing to dominate the baseline on another. This is a substantive reason to report several losses rather than select only the criterion most favorable to the combined method.

The different weights across WVS countries and the U.S. election should not be read as fixed properties of national behavior. They depend on the granularity of the target, the calibration loss, historical-signal construction, the quality of the generated signal, and the stability of relationships across settings. Individual survey responses contain considerable idiosyncratic variation; state vote shares aggregate that variation and may exhibit stronger persistence. Conversely, structural change can reduce the relevance of a formerly strong baseline. Forecast-combination research has long recognized that changing environments can change the value of component forecasts \citep{DieboldPauly1987,Gama2014}. The present results motivate adaptive validation, but do not estimate a general relationship between contextual change and $h$.

The contribution is thus a testable information-integration framework. Our findings highlight the value of Matching-LLM in integrating historical behavioral data with LLM-generated predictions, while showing that the benefits of this combination vary across prediction tasks and evaluation criteria.

\subsection{Factors Affecting Synthetic Prediction Accuracy}

Although Matching-LLM improves predictive performance across several tasks, the results also reveal factors that influence the accuracy of synthetic predictions. Three patterns are particularly relevant. First, a model may give a socially plausible or normatively consistent answer that is poorly calibrated to the population. Near-uniform rejection of theft and bribery is an example: it can appear reasonable while suppressing the variation present in the human sample. Work on synthetic survey data, identity representation, and moral evaluation offers related evidence of uneven fidelity \citep{bisbee2023synthetic,Wang2025,scherrer2024evaluating}. Historical anchoring can introduce observed variation, but it cannot ensure the correct distribution if the historical responses themselves differ from the target.

Second, source errors can align. The Chinese fare-evasion and tax-evasion comparisons show that historical adjustment can amplify an LLM error when the historical mean lies still farther from the current mean in the same direction. The cross-item results show another case: a weight fitted across several attitude domains can place too much emphasis on generated information for items whose historical means are already accurate. These results do not warrant a general conclusion that either historical data or LLM predictions are intrinsically superior.

Third, a method can preserve one aspect of behavior while distorting another. Correct signs of cross-country differences can coexist with strongly attenuated magnitudes. Similar means can coexist with compressed dispersion or undefined correlations. A nearly correct national electoral-vote total can coexist with several incorrect state winners. Recent experimental forecasting research likewise demonstrates that high correlation with observed effects need not imply accurate effect sizes \citep{Ashokkumar2026}. Validation must therefore be tied to the intended substantive use, rather than a single generic notion of human likeness.

Potential explanations for LLM errors include training-data representation, model alignment, persona interpretation, and sensitivity to question formatting \citep{Santurkar2023,Durmus2024,Tjuatja2023Do,Rottger2024}. The current study does not experimentally vary these channels, so it cannot attribute a particular discrepancy to one of them. Similarly, the historical records can reflect nonresponse, measurement error, and social-desirability effects in the original human surveys \citep{phillips1972some}. Observed behavior is a better benchmark for prediction than an assumed idealized response, but it remains a measured signal with its own limitations.

The German application illustrates a broader use of historical information within Matching-LLM. Party-specific multiplicative calibration extends historical grounding to a multiparty prediction task, complementing the common historical weight used in the U.S. application.

\subsection{Limitations and Implications}

Validation depends on the information available for each task. WVS reconstruction and unused-item validation use the same survey wave and respondents, so the latter assesses transfer across questions. The public archives document election forecasts disclosed before the outcomes were known. Historical survey questions and election outcomes may nevertheless have appeared in the LLM's training data, which can affect the interpretation of historical calibration results \citep{KapoorNarayanan2023}.

The findings are conditional on GPT-4o and the selected prompts and generation settings. Their stability across model families and snapshots remains to be tested. Human-grounded agents, behaviorally trained models, and digital-twin datasets provide complementary approaches for future comparisons \citep{Park2026,Binz2025,Toubia2025,Li2025}.

Historical anchoring also depends on demographic overlap and the stability of behavioral relationships. Matching cannot recover unobserved experiences, and ANES weights do not eliminate small state subsamples, nonresponse, or changes in the electorate. Representativeness should therefore be evaluated for the population and prediction task of interest \citep{Stuart2010,Wang2025}.

Uncertainty from historical-donor selection, weight estimation, generation, and dependence among outcomes is only partly represented by the reported intervals and tests. A common country-level weight may also perform differently across attitude domains. The U.S. bootstrap interval summarizes weight sensitivity to state-level sample composition. Repeated temporal validation, alternative matching specifications, and comparisons with conventional statistical predictors would help assess robustness.

The economic implication is that the value of synthetic prediction lies in the additional information it provides per unit of research cost. The framework offers a transparent way to compare that value with an existing historical benchmark. It can support preliminary questionnaire evaluation, scenario exploration, and the prioritization of questions for new data collection, consistent with discussions of LLM use in survey research \citep{jansen2023employing,Bail2024,Sarstedt2024}. However, computationally generating many responses does not create the informational equivalent of recruiting many independent human respondents. Lower generation cost is not sufficient evidence of accuracy or representativeness.

Human data remain necessary for anchoring, evaluation, and detecting change. The study's strongest practical lesson is therefore to assess LLM output together with a credible historical baseline, report the dimensions on which it adds information, and collect new human data where the sources disagree or existing anchors are stale. This approach connects prediction to substantive research needs while preserving the distinction between model diagnostics and causal behavioral explanation \citep{Hofman2021,Shmueli2010}.

\section{Conclusion}\label{sec:conclusion}

This paper asks whether historical human behavior can improve context-conditioned LLM predictions, and whether combining the two sources adds information beyond either one alone. Matching-LLM aligns and integrates historical human behavioral information with LLM-generated predictions at the individual-response and state-share levels, using transparent weights and explicit LLM-only and History-only benchmarks. Its rationale is the conditional value of combining imperfect prediction signals with different error structures.

The WVS evidence shows improved mean reconstruction relative to LLM-only across many items and gains on most questions excluded from weight calibration. It also identifies persistent discrepancies in value-sensitive responses and response heterogeneity. The historical baseline remains better on several comparisons, including all six unused Chinese survey items, while correlation gains relative to History-only occur in the U.S. but not China. These findings establish useful but uneven transfer of historical anchoring, rather than a general advantage of adding LLM output to every historical predictor.

In the 2024 U.S. application, the combination yields a smaller electoral-vote discrepancy than either single source and improves the set of correctly predicted swing states relative to History-only, while leaving total state-winner accuracy unchanged. The German application illustrates a broader historical-calibration extension through party-specific adjustments, with the predicted party distribution and state-leading parties compared with observed outcomes.

The methodological contribution is a historically anchored framework for evaluating and combining synthetic information. Future work should test multiple model snapshots, richer behavioral anchors, alternative matching procedures, and repeated prospective outcomes, with uncertainty appropriate to the intended prediction target. The usefulness of Matching-LLM ultimately depends on demonstrating incremental information against strong baselines while retaining human data as the basis for calibration and validation.

\clearpage
\singlespacing
\setlength\bibsep{0pt}
\bibliographystyle{plainnat}
\bibliography{AEF_ref}

@inproceedings{Aher2022Using,
  title={Using Large Language Models to Simulate Multiple Humans and Replicate Human Subject Studies},
  author={Aher, Gati and Arriaga, Rosa I. and Kalai, Adam Tauman},
  booktitle={Proceedings of the 40th International Conference on Machine Learning},
  series={Proceedings of Machine Learning Research},
  volume={202},
  pages={337--371},
  year={2023},
  publisher={PMLR},
  url={https://proceedings.mlr.press/v202/aher23a.html},
}

@misc{Kim2023AI-Augmented,
  title={{AI}-Augmented Surveys: Leveraging Large Language Models and Surveys for Opinion Prediction},
  author={Kim, Junsol and Lee, Byungkyu},
  year={2026},
  howpublished={arXiv:2305.09620v4},
  doi={10.48550/arXiv.2305.09620},
  url={https://arxiv.org/abs/2305.09620v4},
  note={Revised May 2026; first circulated in 2023},
}

@article{argyle2023out,
  title={Out of one, many: Using language models to simulate human samples},
  author={Argyle, Lisa P and Busby, Ethan C and Fulda, Nancy and Gubler, Joshua R and Rytting, Christopher and Wingate, David},
  journal={Political Analysis},
  volume={31},
  number={3},
  pages={337--351},
  year={2023},
  publisher={Cambridge University Press}
}

@article{bisbee2023synthetic,
  title={Synthetic Replacements for Human Survey Data? The Perils of Large Language Models},
  author={Bisbee, James and Clinton, Joshua D. and Dorff, Cassy and Kenkel, Brenton and Larson, Jennifer M.},
  journal={Political Analysis},
  volume={32},
  number={4},
  pages={401--416},
  year={2024},
  doi={10.1017/pan.2024.5},
}

@article{Tjuatja2023Do,
  title={Do {LLMs} Exhibit Human-like Response Biases? A Case Study in Survey Design},
  author={Tjuatja, Lindia and Chen, Valerie and Wu, Tongshuang and Talwalkar, Ameet and Neubig, Graham},
  journal={Transactions of the Association for Computational Linguistics},
  volume={12},
  pages={1011--1026},
  year={2024},
  doi={10.1162/tacl_a_00685},
  url={https://aclanthology.org/2024.tacl-1.56/},
}

@article{snowberg2021testing,
  title={Testing the waters: Behavior across participant pools},
  author={Snowberg, Erik and Yariv, Leeat},
  journal={American Economic Review},
  volume={111},
  number={2},
  pages={687--719},
  year={2021},
  publisher={American Economic Association 2014 Broadway, Suite 305, Nashville, TN 37203}
}

@article{jansen2023employing,
  title={Employing large language models in survey research},
  author={Jansen, Bernard J and Jung, Soon-gyo and Salminen, Joni},
  journal={Natural Language Processing Journal},
  volume={4},
  pages={100020},
  year={2023},
  publisher={Elsevier}
}

@inproceedings{scherrer2024evaluating,
  title={Evaluating the Moral Beliefs Encoded in {LLMs}},
  author={Scherrer, Nino and Shi, Claudia and Feder, Amir and Blei, David M.},
  booktitle={Advances in Neural Information Processing Systems},
  volume={36},
  year={2023},
  url={https://proceedings.neurips.cc/paper_files/paper/2023/hash/a2cf225ba392627529efef14dc857e22-Abstract-Conference.html},
}

@article{phillips1972some,
  title={Some effects of" social desirability" in survey studies},
  author={Phillips, Derek L and Clancy, Kevin J},
  journal={American journal of sociology},
  volume={77},
  number={5},
  pages={921--940},
  year={1972},
  publisher={University of Chicago Press}
}

@article{BatesGranger1969,
  title={The Combination of Forecasts},
  author={Bates, John M. and Granger, Clive W. J.},
  journal={Operational Research Quarterly},
  volume={20},
  number={4},
  pages={451--468},
  year={1969},
  doi={10.2307/3008764}
}

@article{Clemen1989,
  title={Combining Forecasts: A Review and Annotated Bibliography},
  author={Clemen, Robert T.},
  journal={International Journal of Forecasting},
  volume={5},
  number={4},
  pages={559--583},
  year={1989},
  doi={10.1016/0169-2070(89)90012-5}
}

@article{DieboldPauly1987,
  title={Structural Change and the Combination of Forecasts},
  author={Diebold, Francis X. and Pauly, Peter},
  journal={Journal of Forecasting},
  volume={6},
  number={1},
  pages={21--40},
  year={1987}
}

@techreport{Horton2023,
  title={Large Language Models as Simulated Economic Agents: What Can We Learn from Homo Silicus?},
  author={Horton, John J. and Filippas, Apostolos and Manning, Benjamin S.},
  institution={National Bureau of Economic Research},
  type={Working Paper},
  number={31122},
  year={2023},
  doi={10.3386/w31122}
}

@article{Dillion2023,
  title={Can AI Language Models Replace Human Participants?},
  author={Dillion, Danica and Tandon, Niket and Gu, Yuling and Gray, Kurt},
  journal={Trends in Cognitive Sciences},
  volume={27},
  number={7},
  pages={597--600},
  year={2023},
  doi={10.1016/j.tics.2023.04.008}
}

@inproceedings{Santurkar2023,
  title={Whose Opinions Do Language Models Reflect?},
  author={Santurkar, Shibani and Durmus, Esin and Ladhak, Faisal and Lee, Cinoo and Liang, Percy and Hashimoto, Tatsunori},
  booktitle={Proceedings of the 40th International Conference on Machine Learning},
  series={Proceedings of Machine Learning Research},
  volume={202},
  pages={29971--30004},
  year={2023},
  publisher={PMLR}
}

@article{Wang2025,
  title={Large Language Models That Replace Human Participants Can Harmfully Misportray and Flatten Identity Groups},
  author={Wang, Angelina and Morgenstern, Jamie and Dickerson, John P.},
  journal={Nature Machine Intelligence},
  volume={7},
  number={3},
  pages={400--411},
  year={2025},
  doi={10.1038/s42256-025-00986-z}
}

@article{Bail2024,
  title={Can Generative {AI} Improve Social Science?},
  author={Bail, Christopher A.},
  journal={Proceedings of the National Academy of Sciences},
  volume={121},
  number={21},
  pages={e2314021121},
  year={2024},
  doi={10.1073/pnas.2314021121},
}

@article{Sarstedt2024,
  title={Using Large Language Models to Generate Silicon Samples in Consumer and Marketing Research: Challenges, Opportunities, and Guidelines},
  author={Sarstedt, Marko and Adler, Susanne J. and Rau, Lea and Schmitt, Bernd},
  journal={Psychology \& Marketing},
  volume={41},
  number={6},
  pages={1254--1270},
  year={2024},
  doi={10.1002/mar.21982},
}

@article{Ashokkumar2026,
  title={Large Language Models Can Predict the Results of Social Science Experiments},
  author={Ashokkumar, Ashwini and Hewitt, Luke and Ghezae, Isaias and Willer, Robb},
  journal={Nature},
  volume={656},
  pages={115--122},
  year={2026},
  doi={10.1038/s41586-026-10742-x},
}

@misc{Park2026,
  title={{LLM} Agents Grounded in Self-Reports Enable General-Purpose Simulation of Individuals},
  author={Park, Joon Sung and Zou, Carolyn Q. and Kamphorst, Jonne and Egan, Niles and Shaw, Aaron and Hill, Benjamin Mako and Cai, Carrie and Morris, Meredith Ringel and Liang, Percy and Willer, Robb and Bernstein, Michael S.},
  year={2026},
  howpublished={arXiv:2411.10109v3},
  doi={10.48550/arXiv.2411.10109},
  url={https://arxiv.org/abs/2411.10109v3},
  note={Revised June 2026; first circulated in 2024 as Generative Agent Simulations of 1,000 People},
}

@article{Toubia2025,
  title={Database Report: {Twin-2K-500}: A Data Set for Building Digital Twins of over 2,000 People Based on Their Answers to over 500 Questions},
  author={Toubia, Olivier and Gui, George Z. and Peng, Tianyi and Merlau, Daniel J. and Li, Ang and Chen, Haozhe},
  journal={Marketing Science},
  volume={44},
  number={6},
  pages={1446--1455},
  year={2025},
  doi={10.1287/mksc.2025.0262},
}

@article{Binz2025,
  title={A Foundation Model to Predict and Capture Human Cognition},
  author={Binz, Marcel and Akata, Elif and Bethge, Matthias and Br{\"a}ndle, Franziska and Callaway, Fred and Coda-Forno, Julian and Dayan, Peter and Demircan, Can and Eckstein, Maria K. and {\'E}ltet{\H{o}}, No{\'e}mi and Griffiths, Thomas L. and Haridi, Susanne and Jagadish, Akshay K. and Ji-An, Li and Kipnis, Alexander and Kumar, Sreejan and Ludwig, Tobias and Mathony, Marvin and Mattar, Marcelo and Modirshanechi, Alireza and Nath, Surabhi S. and Peterson, Joshua C. and Rmus, Milena and Russek, Evan M. and Saanum, Tankred and Schubert, Johannes A. and Schulze Buschoff, Luca M. and Singhi, Nishad and Sui, Xin and Thalmann, Mirko and Theis, Fabian J. and Truong, Vuong and Udandarao, Vishaal and Voudouris, Konstantinos and Wilson, Robert and Witte, Kristin and Wu, Shuchen and Wulff, Dirk U. and Xiong, Huadong and Schulz, Eric},
  journal={Nature},
  volume={644},
  pages={1002--1009},
  year={2025},
  doi={10.1038/s41586-025-09215-4},
}

@misc{Durmus2024,
  title={Towards Measuring the Representation of Subjective Global Opinions in Language Models},
  author={Durmus, Esin and Nguyen, Karina and Liao, Thomas I. and Schiefer, Nicholas and Askell, Amanda and Bakhtin, Anton and Chen, Carol and Hatfield-Dodds, Zac and Hernandez, Danny and Joseph, Nicholas and Lovitt, Liane and McCandlish, Sam and Sikder, Orowa and Tamkin, Alex and Thamkul, Janel and Kaplan, Jared and Clark, Jack and Ganguli, Deep},
  year={2024},
  howpublished={arXiv:2306.16388v2},
  doi={10.48550/arXiv.2306.16388},
  url={https://arxiv.org/abs/2306.16388v2},
  note={Revised April 2024; first circulated in 2023},
}

@inproceedings{Li2025,
  title={{LLM} Generated Persona Is a Promise with a Catch},
  author={Li, Ang and Chen, Haozhe and Namkoong, Hongseok and Peng, Tianyi},
  booktitle={Advances in Neural Information Processing Systems},
  volume={38},
  year={2025},
  doi={10.52202/085713-2226},
  url={https://papers.neurips.cc/paper_files/paper/2025/hash/5fc150b75e8b509735564de16bc3527b-Abstract-Position_Paper_Track.html},
  note={Position Paper Track},
}

@inproceedings{Rottger2024,
  title={Political Compass or Spinning Arrow? Towards More Meaningful Evaluations for Values and Opinions in Large Language Models},
  author={R{\"o}ttger, Paul and Hofmann, Valentin and Pyatkin, Valentina and Hinck, Musashi and Kirk, Hannah and Schuetze, Hinrich and Hovy, Dirk},
  booktitle={Proceedings of the 62nd Annual Meeting of the Association for Computational Linguistics (Volume 1: Long Papers)},
  pages={15295--15311},
  year={2024},
  publisher={Association for Computational Linguistics},
  doi={10.18653/v1/2024.acl-long.816},
  url={https://aclanthology.org/2024.acl-long.816/},
}

@article{InglehartBaker2000,
  title={Modernization, Cultural Change, and the Persistence of Traditional Values},
  author={Inglehart, Ronald and Baker, Wayne E.},
  journal={American Sociological Review},
  volume={65},
  number={1},
  pages={19--51},
  year={2000},
  doi={10.1177/000312240006500103},
}

@article{Gama2014,
  title={A Survey on Concept Drift Adaptation},
  author={Gama, Jo{\~a}o and Zliobaite, Indre and Bifet, Albert and Pechenizkiy, Mykola and Bouchachia, Abdelhamid},
  journal={ACM Computing Surveys},
  volume={46},
  number={4},
  pages={44:1--44:37},
  year={2014},
  doi={10.1145/2523813},
}

@article{WangHyndman2023,
  title={Forecast Combinations: An over 50-Year Review},
  author={Wang, Xiaoqian and Hyndman, Rob J. and Li, Feng and Kang, Yanfei},
  journal={International Journal of Forecasting},
  volume={39},
  number={4},
  pages={1518--1547},
  year={2023},
  doi={10.1016/j.ijforecast.2022.11.005},
}

@article{Stuart2010,
  title={Matching Methods for Causal Inference: A Review and a Look Forward},
  author={Stuart, Elizabeth A.},
  journal={Statistical Science},
  volume={25},
  number={1},
  pages={1--21},
  year={2010},
  doi={10.1214/09-STS313},
}

@article{Efron1979,
  title={Bootstrap Methods: Another Look at the Jackknife},
  author={Efron, Bradley},
  journal={The Annals of Statistics},
  volume={7},
  number={1},
  pages={1--26},
  year={1979},
  doi={10.1214/aos/1176344552},
}

@article{Shmueli2010,
  title={To Explain or to Predict?},
  author={Shmueli, Galit},
  journal={Statistical Science},
  volume={25},
  number={3},
  pages={289--310},
  year={2010},
  doi={10.1214/10-STS330},
}

@article{Hofman2021,
  title={Integrating Explanation and Prediction in Computational Social Science},
  author={Hofman, Jake M. and Watts, Duncan J. and Athey, Susan and Garip, Filiz and Griffiths, Thomas L. and Kleinberg, Jon and Margetts, Helen and Mullainathan, Sendhil and Salganik, Matthew J. and Vazire, Simine and Vespignani, Alessandro and Yarkoni, Tal},
  journal={Nature},
  volume={595},
  pages={181--188},
  year={2021},
  doi={10.1038/s41586-021-03659-0},
}

@article{KapoorNarayanan2023,
  title={Leakage and the Reproducibility Crisis in Machine-Learning-Based Science},
  author={Kapoor, Sayash and Narayanan, Arvind},
  journal={Patterns},
  volume={4},
  number={9},
  pages={100804},
  year={2023},
  doi={10.1016/j.patter.2023.100804},
}

@article{GneitingRaftery2007,
  title={Strictly Proper Scoring Rules, Prediction, and Estimation},
  author={Gneiting, Tilmann and Raftery, Adrian E.},
  journal={Journal of the American Statistical Association},
  volume={102},
  number={477},
  pages={359--378},
  year={2007},
  doi={10.1198/016214506000001437},
}

@misc{WVS6,
  title={World Values Survey: Round Six---Country-Pooled Datafile},
  author={Inglehart, Ronald and Haerpfer, Christian and Moreno, Alejandro and Welzel, Christian and Kizilova, Kseniya and Diez-Medrano, Jaime and Lagos, Marta and Norris, Pippa and Ponarin, Eduard and Puranen, Bi},
  year={2018},
  howpublished={JD Systems Institute and World Values Survey Association},
  doi={10.14281/18241.8},
  url={https://www.worldvaluessurvey.org/WVSDocumentationWV6.jsp},
  note={Wave 6 (2010--2014)},
}

@misc{WVS7,
  title={World Values Survey Wave 7 (2017--2022) Cross-National Data-Set},
  author={Haerpfer, Christian and Inglehart, Ronald and Moreno, Alejandro and Welzel, Christian and Kizilova, Kseniya and Diez-Medrano, Jaime and Lagos, Marta and Norris, Pippa and Ponarin, Eduard and Puranen, Bi},
  year={2022},
  howpublished={World Values Survey Association},
  doi={10.14281/18241.18},
  url={https://www.worldvaluessurvey.org/WVSDocumentationWV7.jsp},
}

@misc{ANES2020,
  title={{ANES} 2020 Time Series Study Full Release},
  author={{American National Election Studies}},
  year={2021},
  howpublished={Dataset and documentation},
  url={https://electionstudies.org/data-center/2020-time-series-study/},
}

@misc{GLES2023,
  author={{GLES}},
  year={2023},
  title={{GLES Cross-Section 2021, Pre-Election}},
  howpublished={GESIS, Cologne. ZA7700 Datafile Version 3.1.0},
  doi={10.4232/1.14168},
  url={https://doi.org/10.4232/1.14168}
}

@misc{NARA2024,
  title={2024 Electoral College Results},
  author={{National Archives and Records Administration}},
  year={2024},
  url={https://www.archives.gov/electoral-college/2024},
  note={Official Electoral College results},
}

\section*{Declarations}
\subsection*{Acknowledgements}
The authors declare that no acknowledgments are applicable for this work.

\subsection*{Author contributions}
Shapeng Jiang and Chen Zhang jointly conducted the data analysis. Shapeng Jiang was responsible for data collection and prepared the tables and figures. Chen Zhang conducted the literature review and drafted the manuscript. Lijia Wei developed the study concept, formalized the analytical strategy, acquired funding, and supervised the study. All authors reviewed and approved the final manuscript.

\subsection*{Funding}
This work is supported by the Key Program of the National Natural Science Foundation of China (Grant numbers 72433003).

\subsection*{Competing interests}
The authors declare no competing interests.

\end{document}